\documentclass[alpha-refs]{wiley-article}

\usepackage{siunitx}
\usepackage{cleveref}
\Crefname{figure}{\text{Fig.}}{\text{Figs.}}
\Crefname{equation}{\text{Eq.}}{\text{Eqs.}}
\usepackage{subcaption}
\usepackage[outercaption]{sidecap}
\sidecaptionvpos{figure}{t}
\usepackage[export]{adjustbox}

\papertype{Original Article}
\paperfield{Journal Section}

\title{Maximizing wind farm power output with the helix approach – experimental validation and wake analysis using tomographic PIV}

\author[1]{Daan van der Hoek}
\author[1]{Bert Van den Abbeele}
\author[2]{Carlos Simao Ferreira}
\author[1]{Jan-Willem van Wingerden}

\affil[1]{Delft Center for Systems and Control,  Faculty of Mechanical, Maritime and Materials Engineering (3mE), Delft University of Technology, Delft, The Netherlands}
\affil[2]{Department of Aerodynamics, Wind Energy, Flight Performance and Propulsion, Faculty of Aerospace Engineering, Delft University of Technology, Delft, The Netherlands}

\corraddress{Daan van der Hoek, Delft Center for Systems and Control, Delft University of Technology, 2628 CD Delft, The Netherlands}
\corremail{d.c.vanderhoek@tudelft.nl}

\fundinginfo{This work is part of the Hollandse Kust Noord wind farm innovation program where CrossWind C.V., Shell, Eneco and Siemens Gamesa are teaming up; funding for the PhDs was provided by CrossWind C.V. and Siemens Gamesa.}

\runningauthor{van der Hoek et al.}

\begin{document}

\begin{frontmatter}
\maketitle

\begin{abstract}
\textbf{Abstract}

\noindent Wind farm control can play a key role in reducing the negative impact of wakes on wind turbine power production. The helix approach is a recent innovation in the field of wind farm control, which employs individual blade pitch control to induce a helical velocity profile in a wind turbine wake. This forced meandering of the wake has turned out to be very effective for the recovery of the wake, increasing the power output of downstream turbines by a significant amount. This paper presents a wind tunnel study with two scaled wind turbine models, of which the upstream turbine is operated with the helix approach. We used tomographic particle image velocimetry to study the dynamic behaviour of the wake under influence of the helix excitation. The measured flow fields confirm the wake recovery capabilities of the helix approach compared to normal operation. Additional emphasis is put on the effect of the helix approach on the breakdown of blade tip vortices, a process that plays an important role in re-energizing the wake. Measurements indicate that the breakdown of tip vortices, and the resulting destabilization of the wake is enhanced significantly with the helix approach. Finally, turbine measurements show that the helix approach was able to increase the combined power for this particular two turbine setup by as much as 15\%.

% Please include a maximum of seven keywords
\end{abstract}
\end{frontmatter}

\section{Introduction}
Wakes generated by upstream turbines can negatively impact the performance of downstream turbines and the wind farm as a whole. They contribute to significant losses in energy and increased loading of structural components \citep{Barthelmie2009ModellingOffshore,Barthelmie2013MeteorologicalWakes}. While conventional industry practice takes these negative aspects for granted, there are ways to, at least partially, mitigate wake effects. Wind farm control can play a key role in reducing wake effects by actively influencing the airflow behind turbines. Three wind farm control categories are generally recognized \citep{vanWingerden2020ExpertControl,Meyers2022WindChallenges,Houck2022ReviewTurbines}. \emph{Axial induction control} (AIC) adjusts the induction factor of the turbine either by introducing an offset in the blade pitch angle or running the turbine at a sub-optimal tip speed ratio \citep{Kanev2018,Houck2022ReviewTurbines}. While AIC has proven capable of providing grid services \citep{Aho2012,Fleming2016,Vali2017} and decreasing structural loading \citep{Fleming2016EffectsLoading,VanDerHoek2018ComparisonLoads}, applying it for maximizing power has seen less convincing results \citep{Annoni2016a,Campagnolo2016,Hoek2019,Frederik2020PeriodicExperiments}. The second category is called \emph{wake steering}, and has been shown numerous times to be capable of increasing the power output of wind turbine arrays. By misaligning upstream turbines with respect to the incoming wind and redirecting the wake away from downstream turbines, the latter are able to extract more power and more than overcome the losses incurred by the yaw misalignment \citep{Gebraad2016,Campagnolo2016,Fleming2017b,Fleming2019,Doekemeijer2021FieldItaly}. The last five years have seen an increasing interest in the third category, which is often referred to as \emph{dynamic induction control} (DIC) or \emph{wake mixing}. Methods belonging to this category  make continuous adjustments to the turbines' operating setpoints in order to increase the level of mixing in the wake, and thereby enhance its dissipation. In this paper, we will also focus on a method from this category.

DIC was first applied in large eddy simulations through dynamic variations of a wind turbines' induction factor, and showed large increases in power output \citep{Meyers2015,Munters2017}. A simplified implementation of this concept was introduced by \citet{Munters2018TowardsTurbines}, consisting of a periodic variation in thrust which only depends on an optimal amplitude and frequency. \cite{Frederik2020PeriodicExperiments} tested this method in the wind tunnel on a three turbine array by feeding a sinusoidal pitch reference signal to the upstream turbine. Depending on the frequency and, more importantly, the amplitude of the pitch reference, power gains in the range of 2\,\% were observed. Several studies on the wake recovery mechanism of dynamic induction control were made using both simulations \citep{Ylmaz2018OptimalTurbines,Brown2021AcceleratedInstability,Croce2022AApplications} and wind tunnel experiments \citep{VanDerHoek2022ExperimentalWake}. These studies all showed that dynamic thrust adjustments affect the pairing behavior of subsequent blade tip vortices, a process that is closely associated with re-energizing the wake \citep{Lignarolo2014ExperimentalVelocimetry}. When active, DIC accelerates the vortex pairing, creating a region of high vorticity around which the surrounding flow rolls up. Subsequently, higher energy flow is entrained into the wake.

An extension of the previous method was presented by \citet{Frederik2020TheFarms}, where they used dynamic individual pitch control (DIPC) to induce periodic yaw and tilt moments on the rotor plane. Due to the imposed phase difference in yaw and tilt moments, the location of the thrust force varies periodically around the center of the rotor plane with a near constant radius. This leads to a helical structure in the velocity profile of the wake, for which this technique was dubbed the \emph{helix approach}. By periodically pitching the blades at a slightly lower or higher frequency than the rotor frequency, a clockwise (CW) or counter-clockwise helix (CCW) shape is created. Large eddy simulations with both uniform and turbulent inflow showed that this type of actuation can be very successful in enhancing wake mixing and was able to increase power output of a two turbine array up to 7.5\,\%.

The helix approach was further explored with large eddy simulations in terms of finding the optimal excitation frequency \citep{Muscari2022PhysicsFarms} and amplitude \citep{Korb2021ExploringControl,Taschner2023}. The optimal frequency, usually expressed by the dimensionless Strouhal number ($\mathrm{St}$), was found to be higher compared to DIC with collective pitch actuation. As for the optimal amplitude, \citet{Taschner2023} detected no saturation in the power gain in low turbulence conditions for pitch amplitudes of up to \SI{6}{\degree}, indicating that the power gains will primarily be limited by the level of turbulence and the allowed increase in structural loading caused by the helix approach. Aero-elastic simulations showed significant increases in the fatigue loading of both blades and tower of the actuated turbine, although not as significant as in the case of DIC \citep{Frederik2022OnStrategies,Taschner2023,vanVondelen2023}. The loads of the downstream turbine were also seen to increase by a small amount, but not comparable to the actuated turbine.

Previous work on the helix approach all agrees that the technique is most efficient in the CCW direction when considering power maximization. A similar comparison between the two helix counterparts was also made by \citet{CoqueletThesis}, who simulated the performance of the helix approach using a vortex-particle-mesh code. These simulations once more showed the superiority of the CCW helix over the CW version. However, the simulation results also indicated that the potential power gain of the helix decreases significantly as the resolution of the simulations increases. This suggests that previously reported gains obtained through large eddy simulations might be largely overestimated, and in case of the CW helix might not even be present.

The work from \citet{CoqueletThesis} highlights the importance of validating wind farm control methods in real life scenarios, starting with wind tunnel experiments and followed by field testing. This standpoint is reinforced by experts in both academia and industry, who rank a lack of validation as the primary barrier for implementing wind farm control methods on commercial wind farms \citep{vanWingerden2020ExpertControl}. A first step in the validation of the helix approach was taken by \citet{FlorianHeckmeier}, who successfully implemented it in the wind tunnel on a two turbine array. A multi-hole pressure probe was used to measure the velocity profile of a horizontal slice of the wake, indicating increased wake recovery. While these measurements allowed for a phase-locked representation of the wake based on the azimuthal position of the blades, the true transient behaviour of the helix approach was not observed. 

In this paper, we expand upon the experiments done by \citet{FlorianHeckmeier}. Using a similar two wind turbine setup, a full three-dimensional quasi-time-resolved characterization of the wake is obtained using particle image velocimetry (PIV). Individual pitch control can be problematic for small scale turbines, both due to limited space for actuators as well as higher pitch rates. To overcome this challenge, the turbine was equipped with a swashplate that enables cyclic pitch adjustments. This paper will contribute in a major way to the existing literature in two ways. First, the work is seen as another important validation of the helix approach using an alternative turbine model and in a different wind tunnel. A contribution that will hopefully help pave the way to future field testing. Second, the PIV measurements allow for an extensive analysis of the dynamic behaviour in the wake as a result of the helix approach, shedding more light on its working principles. 

\section{Experiment setup}\label{sec:experiment_setup}

\subsection{Wind turbine model}
For this experiment we used two models of the MoWiTO-0.6 wind turbine developed by the University of Oldenburg \citep{Schottler2016DesignStudies}. The turbine has a rotor diameter of $D=\SI{0.58}{\meter}$. The rotor speed can be controlled by regulating the generator torque. The standard turbine model is equipped with a single stepper motor that pitches the blades collectively. To achieve the necessary individual pitch action required for replicating the helix approach, some adjustments to the turbine were necessary. Due to the dimensions of the turbine, using three motors to drive each pitch bearing directly was not a feasible option. The solution was to implement a swashplate mechanism that is commonly used in helicopters. 

\begin{SCfigure}
     \includegraphics[width=0.7\linewidth]{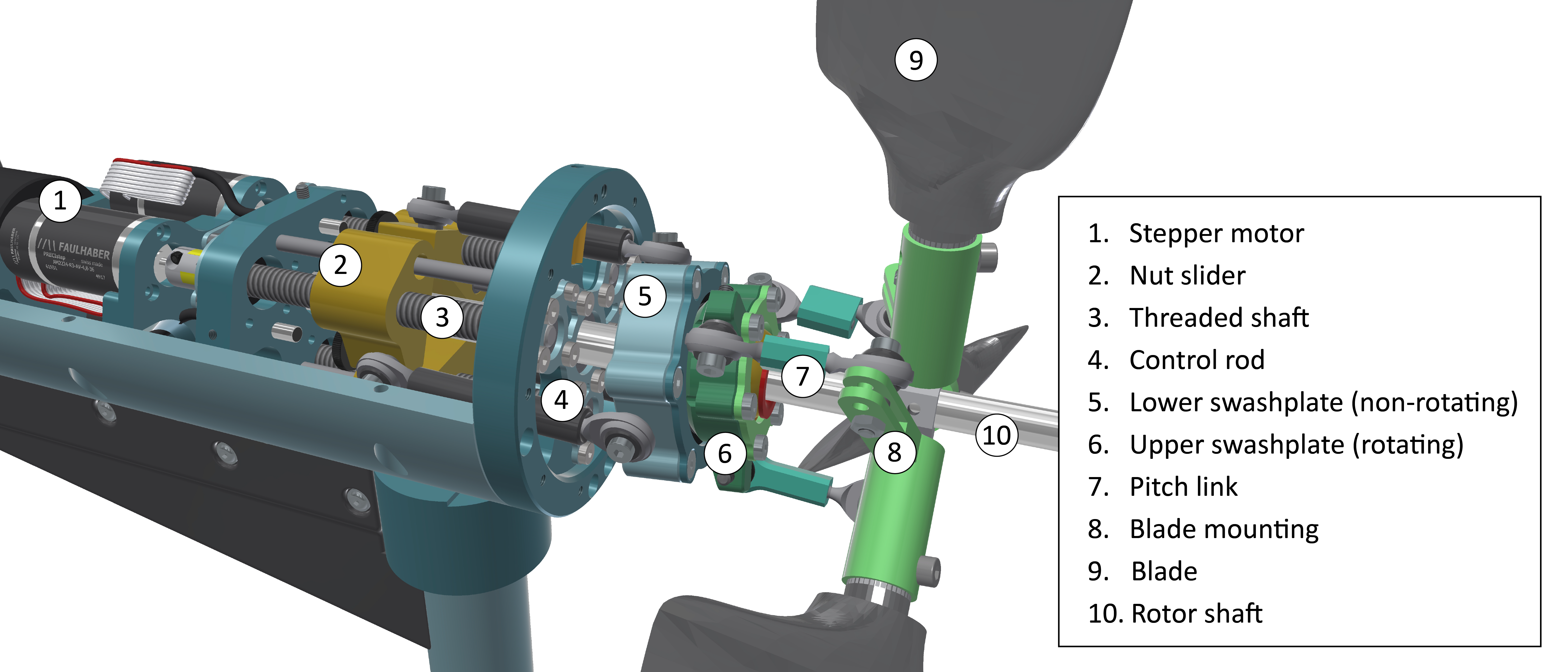}
    \caption{Overview of the swashplate design listing all the involved components. The stepper motors are used to move nut sliders over a threaded shaft, which are in turn connected to the lower swashplate. The upper swashplate assumes the same orientation as the lower swashplate, resulting in different pitch angles for each blade.}
    \label{fig:swashplate}
\end{SCfigure}

A visualization of the swashplate design is given in \Cref{fig:swashplate}. The swashplate consists of a non-rotating part that is actuated by three stepper motors driving three control rods along threaded shafts. The rotating component of the swashplate is connected to the blade mounting mechanism using three pitch links. Both swashplate components are fitted on the rotor shaft using spherical bearings, allowing sufficient range of motion. By tilting the lower swashplate to an arbitrary position, the upper swashplate is forced to assume the same position, which in turn leads to a different pitch setting for each of the three blades. When the rotor is spinning, this will automatically result in a sinusoidal adjustment of the pitch angle of each blade over a single revolution.

The deformation of the tower in the fore-aft direction was measured using a set of strain gauges applied to the bottom of the tower. The strain gauges were connected in a full Wheatstone bridge configuration, with strain expressed as a single voltage measurement. Prior to testing, the strain gauges were calibrated in order to obtain an expression of the bending moment at the tower base as a function of voltage. 

The control inputs for the stepper motors and generator torque 
were sent through a \textit{dSPACE MicroLabBox} real-time system, which was also connected to a computer. A \textsc{Simulink} model of the wind turbine was compiled on the \textit{dSPACE} system, which was called to send and receive data using a MATLAB script. Communication took place at a frequency of \SI{2}{\kilo\Hz}, where besides sending the control input signals, sensor data was acquired of the stepper motor encoders, generator current, generator speed, rotor azimuth position, and tower bottom strain.  
At the start of the experimental campaign, the optimal operating conditions were determined empirically for both turbines. The power coefficient was determined using the measured generator current, which was multiplied by a generator specific torque constant and the rotor speed to obtain electrical power \citep{Schottler2016DesignStudies}. The power coefficient was seen reaching its maximum value of $C_{P_e}\approx0.27$ with a blade pitch angle of $\theta=\SI{10}{\degree}$, and tip-speed ratio of $\lambda=\Omega R/U_\infty=5$. Here, $\Omega$ refers to the rotor speed, $R$ the rotor radius, and $U_\infty$ the free stream inflow velocity. The definition of the pitch angle is based on assigning a \SI{0}{\degree} pitch angle to the most outward position the blades could be pitched to (i.e., pitched fully into the incoming flow). The generator torque of both turbines was regulated in two different ways. For the upstream turbine, we wanted to avoid any effect a varying rotor speed could have on the thrust force. Hence, a straightforward proportional-integral feedback controller was implemented to regulate the torque based on the measured rotor speed, ensuring a near constant rotor speed, and consequently a constant $\lambda$, during all measurements. For the downstream turbine, the incoming flow velocity was unknown, meaning that the previous strategy could not be implemented. In order to optimize the power extraction for the downstream turbine, the torque was regulated based on the measured rotor speed with the widely used $K\Omega^2$-control law \citep{Bossanyi2000TheTurbines}.

\subsection{Implementing the helix approach}
\begin{figure}[t!]
    \centering
    \includegraphics[scale = 1]{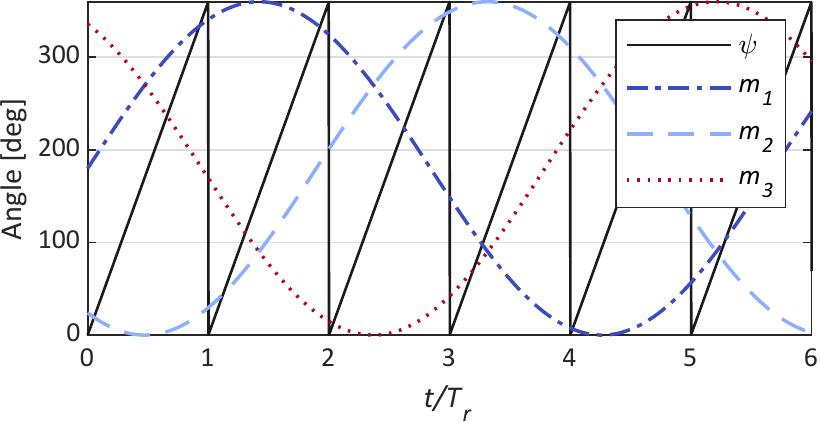}\hfill
    \includegraphics[scale = 1]{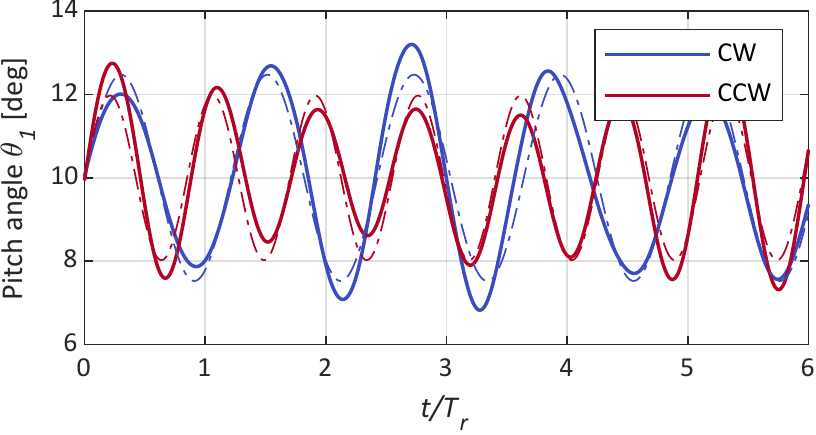}
    \caption{The left figure shows an example of CW actuation of the three stepper motors ($m_1$, $m_2$, and $m_3$) used for controlling the motion of the swashplate. The black line indicates the rotor azimuth  position $\psi$. The right figure shows a comparison between a CW and CCW motion of the swashplate and the resulting pitch angles for one of the blades. The dashed lines show the ideal pitch actuation based on the average pitch amplitude of both helix cases. In both figures, time $t$ is normalized by the turbine rotation period $T_r$}
    \label{fig:pitchAngles}
\end{figure}

With the adjusted turbine model presented in the previous section, actuating the turbine to recreate the helix is relatively straightforward. The original helix control strategy used the multi-blade coordinate (MBC) transformation \citep{Bir2008Multi-bladeAnalysis} to transform the varying tilt and yaw moments to inputs for the pitch controller. The swashplate can be viewed as a mechanical counterpart to the MBC transformation, meaning that we only have to supply the desired frequency and amplitude of the swashplate motion as inputs to our wind turbine. Since the swashplate actuators are spaced \SI{120}{\degree} apart, moving the swashplate periodically in a CW or CCW motion requires applying a sinusoidal reference to the second and third stepper motors with offsets of $\pm\SI{120}{\degree}$ and $\pm\SI{240}{\degree}$. In this case, a positive offset results in a clockwise movement of the swashplate when viewing the turbine from the front. An example of the control inputs is provided on the left side of \Cref{fig:pitchAngles}. 

An initial study into the optimal amplitude and frequency of the swashplate motion was performed prior to the PIV measurements. By varying the amplitude and frequency, and recording the combined power of the two turbines, the settings were found that resulted in optimal power extraction at a wind speed of $U_{\infty}=\SI{5}{\meter\per\second}$. The highest power gain for the CW case was found with an amplitude of 0.5 revolutions (\SI{180}{\degree}) of the stepper motors, while an amplitude of 0.6 revolutions (\SI{216}{\degree}) showed the largest gain for the CCW case. It has to be noted that the latter was the highest amplitude the stepper motors were able to reach at this particular wind speed, without faults being introduced in the tracking of the reference signals. The optimal excitation frequency $f_e$, expressed by the Strouhal
number $\mathrm{St} = f_e D/U_{\infty}$, was determined to be at $\mathrm{St}=0.34$ and $\mathrm{St}=0.28$ for the CW and CCW helix cases, respectively. The resulting excitation frequencies are very close to the optimal frequency of $f_e = 0.18f_r$ reported by \citet{FlorianHeckmeier}. Here, $f_r$ represents the rotational frequency of the turbine. For the PIV measurements we therefore decided to implement both helix cases at these optimal frequencies, while using an amplitude of \SI{180}{\degree} for the stepper motor actuation to compare both methods. Additional measurements were also taken for the CCW case with an amplitude of \SI{216}{\degree}.

Measurements of the actual blade pitch angles were taken after the experimental campaign. By fitting a potentiometer directly to each of the blade mounting mechanisms seen in \Cref{fig:swashplate}, the pitch angles were recorded for different azimuth positions and orientations of the swashplate. The evolution of the pitch angle of a single blade over time is presented on the right hand side of \Cref{fig:pitchAngles} for one amplitude of the stepper motors. This figure shows the difference in pitch rates of the two methods. As the swashplate is moving in the CW direction, similar to the rotational direction of the rotor, the blades are not able to complete a full pitch cycle over a single rotation of the rotor. The opposite is true for the CCW case, resulting in a pitch frequency that is slightly higher than the rotational frequency. Apart from the differences in pitch rate, one can also observe that for both methods the amplitude varies over time. This is the result of some compromises that were made when designing the swashplate to allow sufficient freedom of movement. The dashed lines in the figure show the ideal pitch angles based on the average amplitudes that were measured using the potentiometer. An unexpected property of the swashplate design is that for a single motor amplitude, the resulting pitch amplitudes for the CW and CCW direction are not equal (i.e., the swashplate experiences more/less freedom of movement depending on the direction it is moving). The difference in average pitch amplitude is something that should be considered when analyzing the measurement results in \Cref{sec:results}.

\begin{SCfigure}[][tb]
    \caption{Fixed frame tilt and yaw angles for the CCW helix implementation as function of time normalized by the excitation period $T_e=1/f_e$. The dashed lines show the ideal tilt and yaw angles, in the case of a constant amplitude of the pitch angles in the rotating reference frame.}
    \label{fig:MBCtransform}
    \includegraphics[scale = 1]{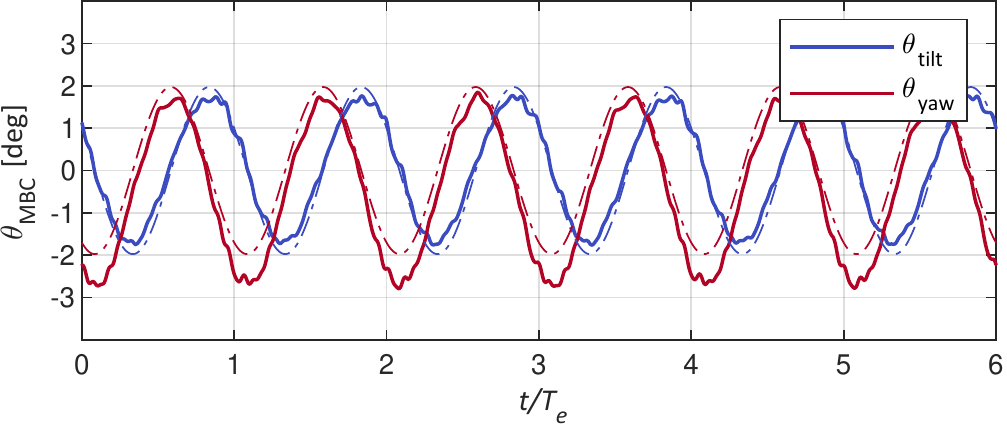}
\end{SCfigure}

The variation of the blade pitch amplitude over time will also have an impact on the quality of the yaw and tilt moments that are being exerted on the incoming flow. We can observe the extent of this variation using the MBC transformation \citep{Bir2008Multi-bladeAnalysis}
\begin{equation}
    \left[\begin{matrix}
        \theta_0(t)\\ 
        \theta_\text{tilt}(t)\\
        \theta_\text{yaw}(t)
    \end{matrix}\right] = \frac{2}{3}\left[\begin{matrix}
        0.5 & 0.5 & 0.5\\ 
        \cos(\psi) & \cos(\psi+\frac{2}{3}\pi) & \cos(\psi+\frac{4}{3}\pi)\\
        \sin(\psi) & \sin(\psi+\frac{2}{3}\pi) &\sin(\psi+\frac{4}{3}\pi)
    \end{matrix}\right] \left[\begin{matrix}
        \theta_1(t)\\ 
        \theta_2(t)\\
        \theta_3(t)
    \end{matrix}\right],
\end{equation}
where $\theta_0$, $\theta_\text{tilt}$, and $\theta_\text{yaw}$ represent the fixed reference frame collective, tilt and yaw angles. Figure~\ref{fig:MBCtransform} presents the transformed tilt and yaw angles for the CCW helix based on the measured pitch angles. The measured angles are compared to the ideal tilt and yaw reference angles, shown by the dashed lines, in case of a constant amplitude on the blade pitch inputs. From the figure it is evident that the swashplate is able to move the pitch angles in such a way to obtain a $\SI{90}{\degree}$ offset between the tilt and yaw angles. While the tilt angle shows a good fit to ideal reference, it seems that some imperfections in the swashplate design result in a small bias for the yaw angle. Based on this, a small warping of the helix velocity profile can be expected.

\subsection{Wind tunnel experiment design}
The experiment was carried out at TU Delft's Open Jet Facility. The wind tunnel consists of an octagonal shaped open jet with dimensions of $\SI{2.85}{\meter}$ by $\SI{2.85}{\meter}$, which is run in a closed circuit. The flow is passed through a radiator system to keep the flow temperature near constant. Before exiting the jet, the flow is contracted by a 3:1 ratio, reaching velocities of up to $\SI{35}{\meter\per\second}$ at turbulence intensities (TI) ranging between 0.5--2\,\% \citep{Lignarolo2015Tip-vortexWakes}. During all measurements of the experimental campaign, the wind tunnel was operated at a constant velocity of $U_\infty=\SI{5}{\meter\per\second}$. 

\begin{figure}[t]
    \centering
    \includegraphics[width=1\linewidth,trim = {0cm, 0.0cm, 0cm, 0.0cm}, clip]{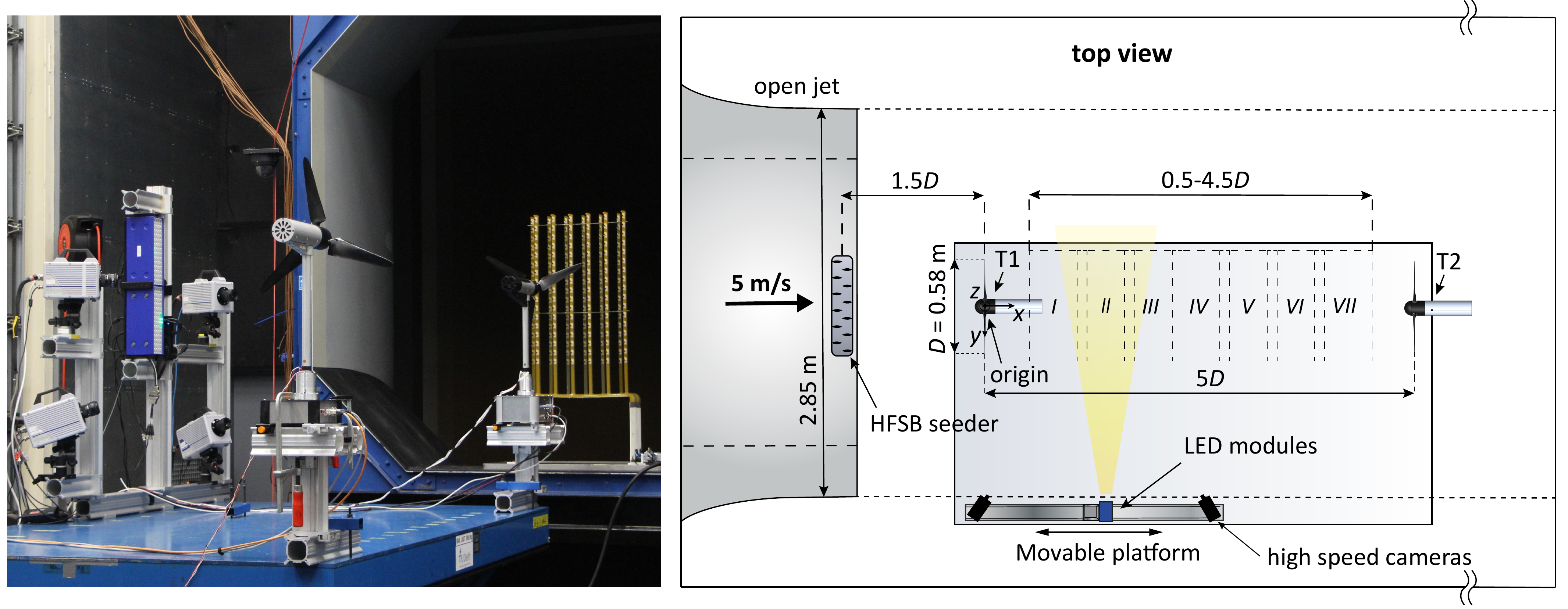}
    \caption{Overview of the experimental setup, highlighting the different components and the measured sections of the wake. The field of views indicated by the numbers \textit{I--VII} are centered at distances of $1:0.5:4\,D$ behind the upstream wind turbine. The cameras and LEDs were mounted on a frame that could be moved along the platform accordingly.}
    \label{fig:Experimental_setup}
\end{figure}

The two MoWiTO-0.6 turbines were placed on a large platform, set apart by a distance of 5 rotor diameters ($D$), similar to other studies on the helix approach \citep{Frederik2020TheFarms,FlorianHeckmeier}. The remaining measurement equipment is illustrated in \Cref{fig:Experimental_setup}. A seeding rake was placed at the outlet of the open jet, releasing helium-filled soap bubbles (HFSBs) that were used as flow tracers over an area \SI{0.6}{\meter} wide by \SI{0.9}{\meter} high \citep{Scarano2015OnExperiments}. On their way downstream, the flow tracers were illuminated by two \textit{LaVision LED Flashlights}. The resulting reflections were recorded by a tomographic PIV system consisting of four \textit{Photron FASTCAM SA1.1} high-speed cameras, operating with a framerate of \SI{500}{frames\per\second} and at a resolution of 1024 by 1024 pixels. Each recording acquired 5,000 frames, resulting in approximately 24 cycles of the helix approach. Synchonization between cameras and LEDs was arranged using a \textit{LaVision PTU-X} timing device. The measurement domain was divided into seven smaller volumes of size \SI{0.4}{} x \SI{0.7}{} x \SI{0.7}{m}$^3$, which were covered by moving the construction on which the cameras and LEDs were mounted. The measurement volumes overlapped in the streamwise direction ($x$), resulting in a total volume of \SI{2.1}{} x \SI{0.7}{} x \SI{0.7}{m}$^3$ and covering a portion of the turbine wake ranging from $0.7$--$4.3\,D$. An overview of all the test cases that were measured is presented \Cref{tab:test_cases}.

\begin{table}[bt]
    \caption{All test cases that were considered for the PIV measurements, indicating the Strouhal number, average pitch amplitude $A_\theta$, the total number of frames that were recorded for each measurement volume, and the total number of helix cycles that were completed during the measurements.}
    \label{tab:test_cases}
    \begin{threeparttable}
        \begin{tabular}{lcccc}
         \headrow
         \thead{Test case\hspace{1cm}} & \thead{\hspace{0.3cm}Strouhal nr.\hspace{0.3cm}} & \thead{\hspace{0.3cm}$\boldsymbol A_\theta$\hspace{0.3cm}} & \thead{\hspace{0.3cm}\#
         frames\hspace{0.3cm}} & \thead{\hspace{0.3cm}\# helix cycles\hspace{0.3cm}}\\
         Baseline (greedy) & n.a. & n.a. & 10,000 & n.a. \\
         Helix CW & 0.34 & \SI{2.5}{\degree} & 15,000 & 82\\
         Helix CCW & 0.28 & \SI{2.0}{\degree} & 15,000 & 72 \\
         Helix CCW & 0.28 & \SI{2.4}{\degree} & 5,000 & 24\\
         \hline  % Please only put a hline at the end of the table
         \end{tabular}
        %\begin{tablenotes}
        %\item $A_\theta$ indicates the average blade pitch amplitude for the helix approach.
        %\end{tablenotes}
    \end{threeparttable}
 \end{table}
 
The PIV data was processed using \textit{LaVision's DaVis 10} software. A high-pass filter was applied to the raw data to reduce any background illumination \citep{Sciacchitano2014EliminationFilter}. A geometrical mapping of the camera positions relative to the measurement domain was obtained using a calibration plate, followed by a volume self-calibration \citep{Wieneke2008VolumeVelocimetry} to reduce the residual measurement errors below 0.1 pixels. Next, we reconstructed the flow tracers with the Shake-The-Box algorithm \citep{Schanz2016Shake-The-Box:Densities}, resulting in approximately 20,000 identified particle tracks per time step. The final processing step consisted of mapping the particle track data onto a Cartesian grid by spatially averaging over small cells. For this binning procedure, we used $\SI{40}{}$ x $\SI{40}{}$ x $\SI{40}{\milli\meter}^3$ cells with a Gaussian weighing function. An overlap of 75\,\% was selected to obtain a grid spacing of $\SI{10}{\milli\meter}$ in the resulting velocity fields. 

The measurement setup has many similarities to the one used in \citep{VanDerHoek2022ExperimentalWake}, where we measured the effect of periodic dynamic induction control with collective pitch on a turbine wake. Some improvements that were made for the current experiment compared to the previous setup consisted of: 1) A movable construction for the cameras, opposed to moving the turbine with respect to the jet outlet, which was seen to have some impact in the turbine inflow conditions. 2) Adding a fourth camera and increasing the tomographic angle (i.e., the subtended angle of all cameras with respect to the measurement domain). This resulted in improved reconstruction of the velocity fields and an increased field of view (FOV). 3) Multiple repetitions of the same control action were recorded for improved convergence of the velocitiy fields. 4) The distance to the HFSB seeder was increased somewhat to reduce its impact on the inflow conditions. 5) The total length of the measurement domain was increased to capture a larger portion of the wake. 

\begin{figure}[b!]
    \centering
    \includegraphics[scale=1,trim = {0, 0, 0.cm, 0}, clip, valign=t]{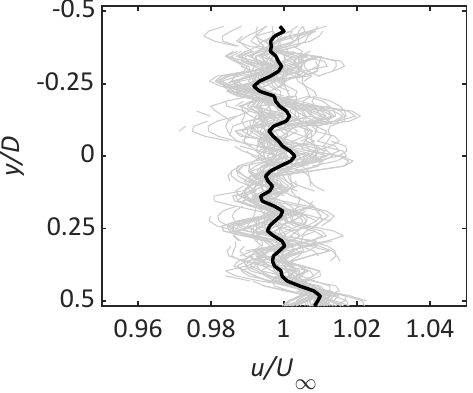}
    \includegraphics[scale=1,trim = {0.97cm, 0, 0, 0}, clip, valign=t]{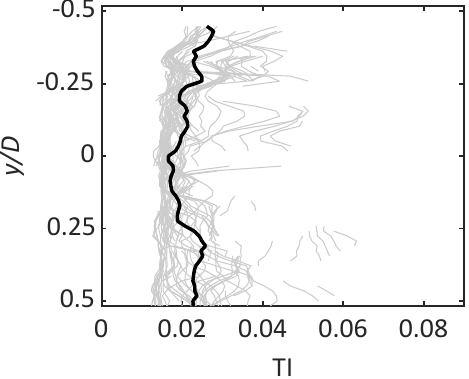}
    \hfill
    \includegraphics[trim = {0, 0, 0, 0}, clip, scale=1,valign=t]{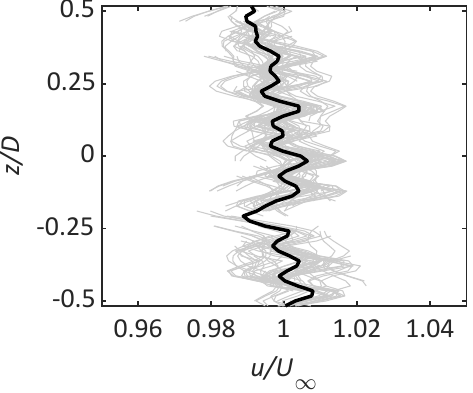}
    \includegraphics[trim = {0.97cm, 0, 0, 0}, clip, scale=1,valign=t]{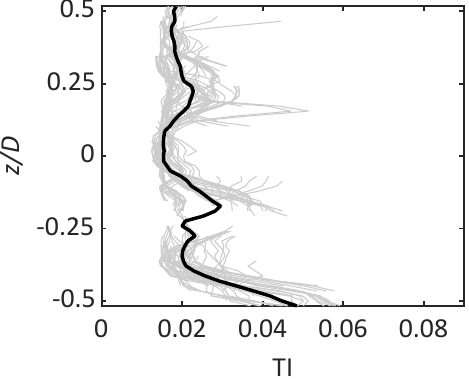}
    \caption{Horizontal ($y$) and vertical ($z$) profiles of the inflow obtained without the presence of a turbine. The grey lines indicate the flow profiles at a single height or width, while the black lines represent the flow profiles averaged over all heights or widths. Note that not all parts of the measurement volume were seeded sufficiently with flow tracers, due to some blocked channels of the seeding rake.}
    \label{fig:freestream_inflow}
\end{figure}

Prior to the helix experiments, we measured the flow without any turbines at a distance of $2.5\,D$ downstream of the seeding rake to get an approximation of the inflow conditions at the upstream turbine (T1). The average measured wind speed and TI profiles at a hypothetical rotor plane are given in \Cref{fig:freestream_inflow}. Both the horizontal and vertical velocity profiles are seen to be centered around the specified inflow speed of $U_\infty=\SI{5}{\meter\per\second}$. The TI in both directions slightly exceeded the earlier specified level of 2\,\%, indicating only a marginal increase in turbulence at the inflow of the turbine as a result of the seeding rake. 

\section{Results} \label{sec:results}
The flow fields that are presented in this section only consider the first three test cases from \Cref{tab:test_cases}. The fourth test case consisting of the CCW helix with larger pitch amplitude had a smaller amount of measurement data, meaning that the resulting flow fields (primarily phase-averaged) are of somewhat lesser quality. For this reason, we decided to omit the flow fields in the paper, apart from some additional analysis of the power performance. Please note when comparing the remaining flow fields of the helix approach, that there is a small difference in average pitch amplitude between CW and CCW actuation, making a one on one comparison more difficult.

\subsection{Time-averaged flow fields}
\begin{figure}[t!]
\centering
    \includegraphics[trim = {0, 0.7cm, 0, 0}, clip,scale=1]{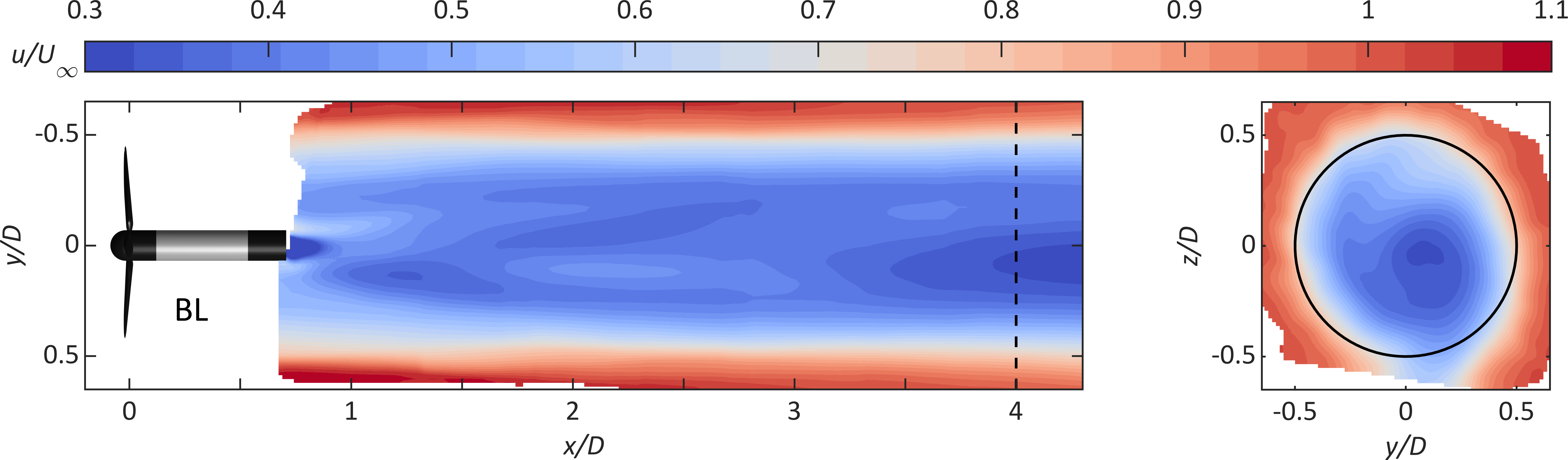}\\
    \includegraphics[trim = {0, 0.7cm, 0, 0.92cm}, clip,scale=1]{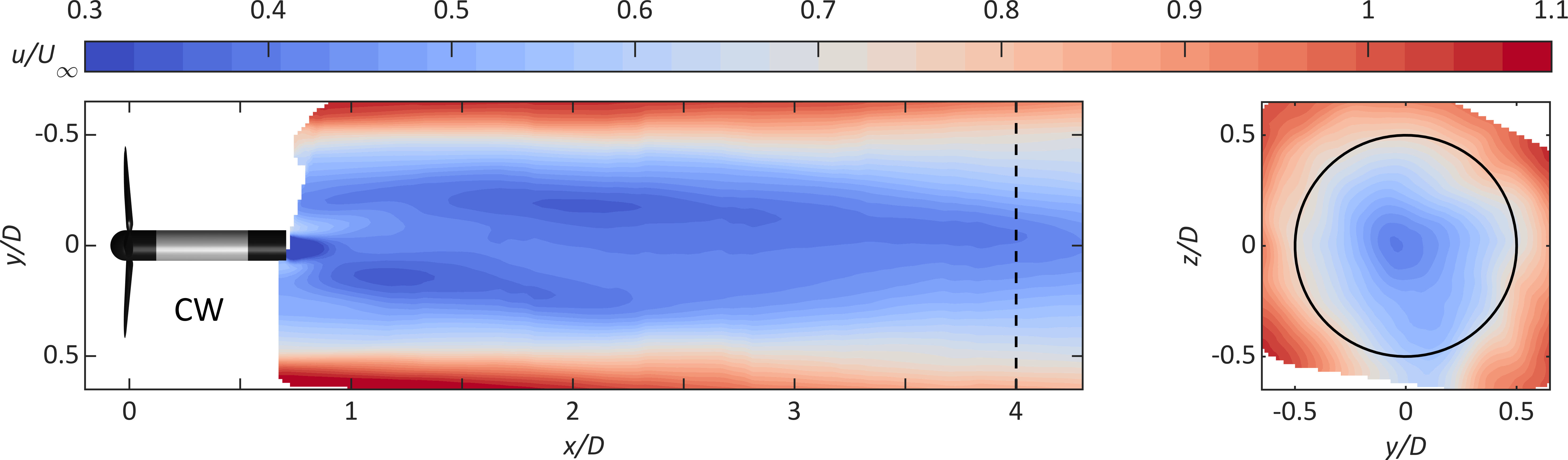}\\
    \includegraphics[trim = {0, 0, 0, 0.92cm}, clip,scale=1]{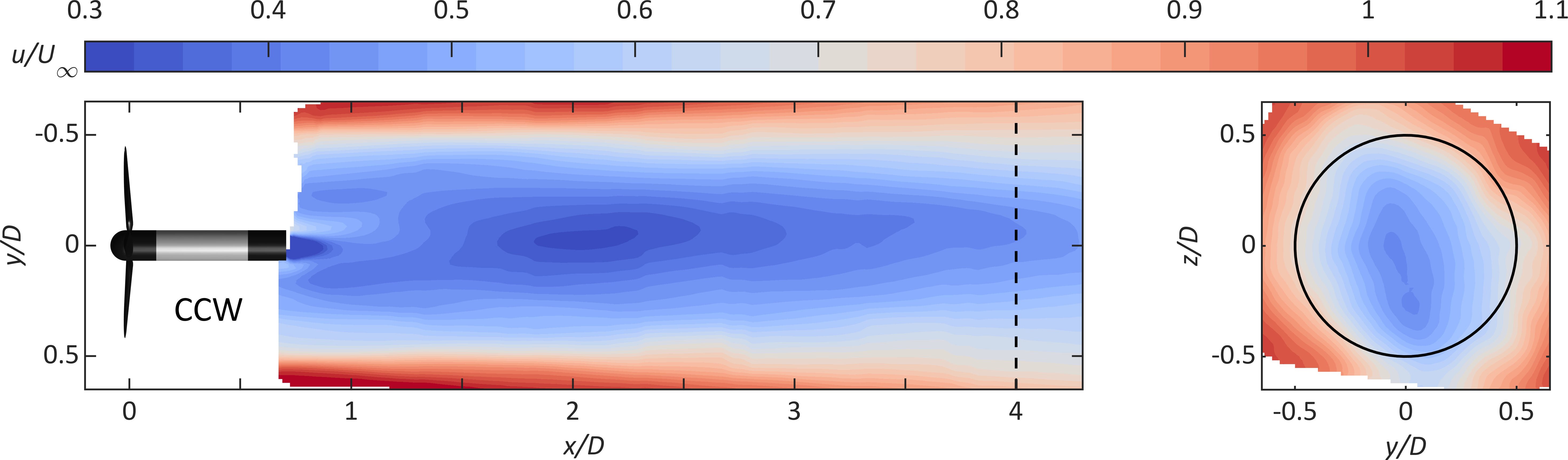}
    \caption{Normalized streamwise velocity contours $u/U_\infty$ reconstructed from the time-averaged particle data. The velocity contours at $z/D = 0$ and $x/D=4$ compare baseline operation with the CW and CCW helix implementation.}
    \label{fig:TA_vel_fields}
\end{figure}

The binning procedure described in \Cref{sec:experiment_setup} was initially applied to all particle data at once, providing the time-averaged flow fields. The streamwise velocity fields of the turbine operating under different control strategies are shown in \Cref{fig:TA_vel_fields}. For the baseline case, where the turbine is operated using \emph{greedy} control, the horizontal cross section of the turbine wake remains relatively constant. This indicates that there is little recovery in the portion of the wake that was measured. Both helix cases show a clear improvement in wake recovery, marked by a larger wake expansion, but a narrower wake deficit profile. This is most obvious from the vertical cross section of the wake at $x/D=4$. Here, we also observe some differences in the shape of the wake between the CW and CCW helix, where the latter seems to be stretched more in the vertical direction. 

\begin{figure}[t!]
    \centering
    \includegraphics[trim = {0, 0.65cm, 0, 0}, clip,scale=1]{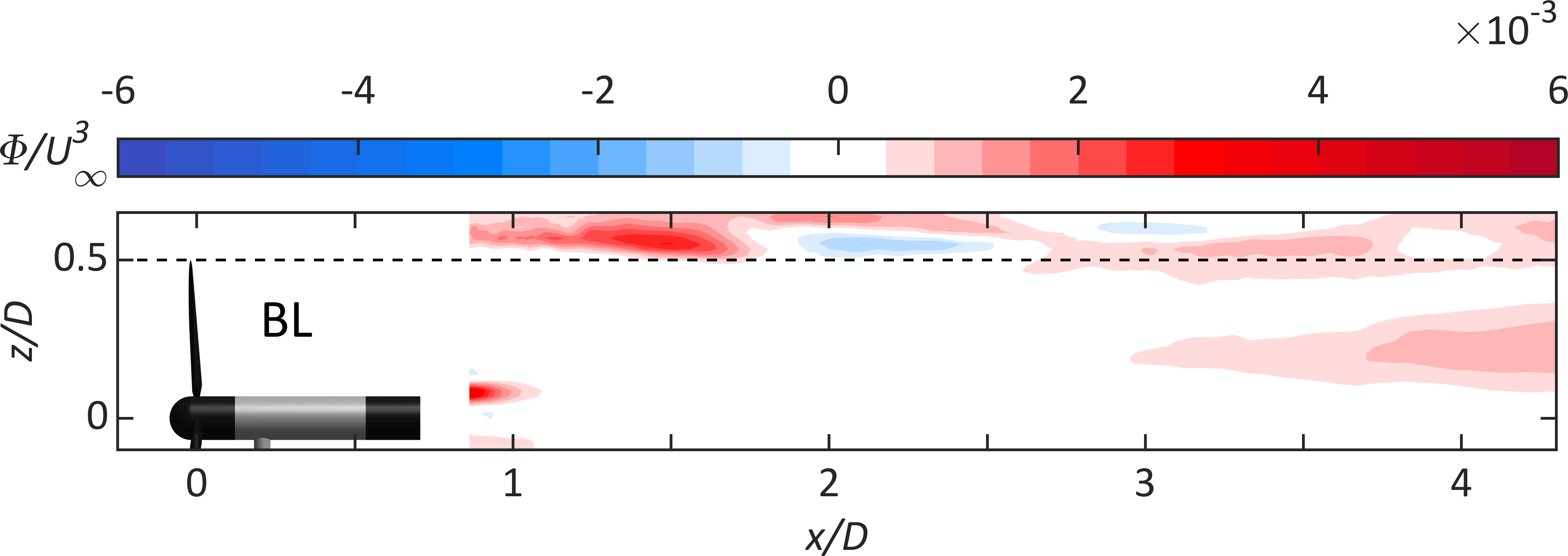}\\
    \includegraphics[trim = {0, 0.65cm, 0, 1.36cm}, clip,scale=1]{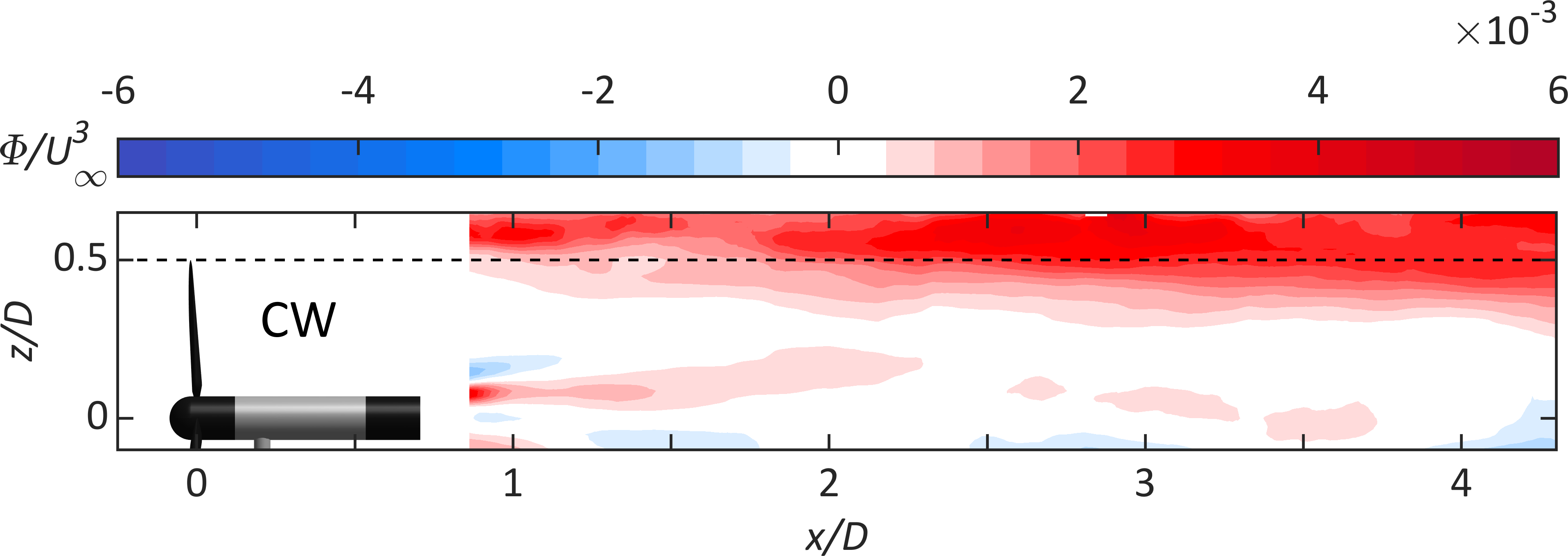}\\
    \includegraphics[trim = {0, 0cm, 0, 1.36cm}, clip,scale=1]{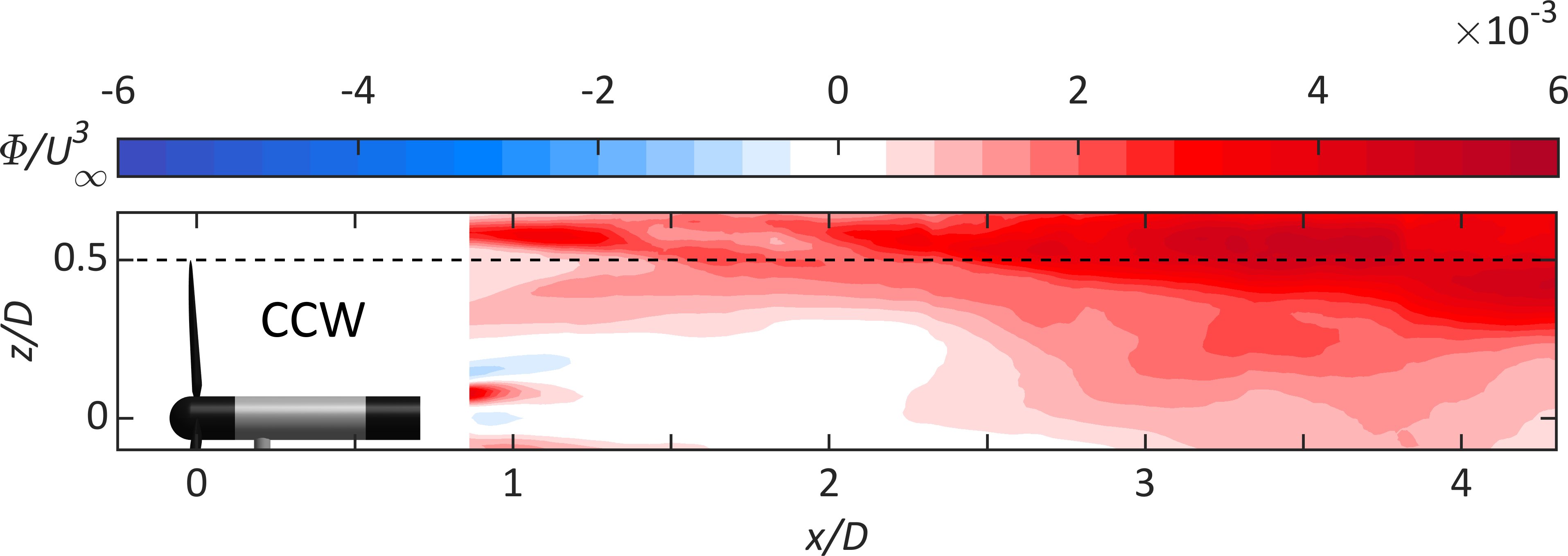}
     \caption{Vertical flow field ($y/D=0$) of the normalized flux of mean kinetic energy into the wake region. A negative definition of the flux is taken, meaning that red areas indicate flux of kinetic energy in the negative $z$ direction (downward), and vice versa for blue areas. We are primarily interested in the flux crossing the boundary of the wake, marked by the dashed line. The figure shows a comparison between baseline operation, CW helix, and CCW helix.}
    \label{fig:Ta_mke_fields}
\end{figure}

An indication of the level of wake recovery or \emph{entrainment} is given by the flux of mean kinetic energy across the boundary layer of the wake. Starting from the mean kinetic energy equation, \citet{Cal2010ExperimentalLayer} showed that the flux of mean kinetic energy due to Reynolds shear stresses is the dominant factor for the entrainment of energy into a turbine wake. An expression for the flux of mean kinetic energy can be obtained using the triple decomposition for a velocity component $u$ \citep{Reynolds1972TheExperiments}:
\begin{equation}
    u = \bar u + u',
\end{equation}
with $\bar u$ the time-averaged velocity component, and $u'$ the fluctuation of the velocity component. The latter term can be further decomposed as
\begin{equation}
    u' = \tilde u + u_s,
\end{equation}
where $\tilde u$ and $u_s$ represent periodic and random fluctuations in the flow, respectively. The flux of mean kinetic energy due to shear stress in the vertical plane is subsequently given by
\begin{equation}
    \Phi = -\overline{u'w'}\bar{u}.
\end{equation}
The term $\overline{u'w'}$ indicates the time-averaged Reynolds shear stress in the vertical plane, which is computed as 
\begin{equation}
\label{eq:reynolds_stress}
    \overline{u'w'} = \frac{\sum_{k=1}^N\left(u(t_k)-\bar{u}\right)\left(w(t_k)-\bar{w}\right)}{N},
\end{equation}
with $u(t_k)$ and $w(t_k)$ the in-plane velocity components at time instant $t_k$, for a number of samples $N$. The improved wake mixing achieved with both helix methods can be seen in \Cref{fig:Ta_mke_fields}, which shows the entrainment or flux of mean kinetic energy into the region of the wake outlined by $z/D=0.5$. Looking at the baseline case, we initially see little entrainment across the wake boundary layer. Around a distance of $x/D=2$, one can observe a clear sign flip in the flux of mean kinetic energy. This sign flip was identified as the location where the leapfrogging or pairing of trailing blade tip vortices occurs \citep{Lignarolo2014ExperimentalVelocimetry}. Following the completion of this leapfrogging motion, an increase in kinetic energy entrainment into the wake was seen. Observing this phenomenon, \citet{Lignarolo2014ExperimentalVelocimetry} concluded that leapfrogging is closely related to the re-energizing of the wake, and that accelerated breakdown of the tip-vortices will result in earlier and thus an increased recovery of wake velocity. 

For both helix cases, \Cref{fig:Ta_mke_fields} shows large increases in entrainment, starting earlier than the baseline case and rising even more after $x/D=2$. Unlike the baseline case, we do not see a clear sign flip in the entrainment indicating the leapfrogging location. This is the combined result of a varying leapfrogging location, and the forced meandering of the wake as a result of the periodic tilt and yaw moments exerted at the rotor, leading to a more diffused entrainment profile. The variation in leapfrogging location was also observed for DIC by collective blade pitch in both simulations \citep{Croce2022AApplications} and a wind tunnel study \citep{VanDerHoek2022ExperimentalWake}. In case of DIC, the periodic thrust variations lead to differences in the strength and expansion of the tip vortices. This increases the reciprocal influence of the helical vortices, triggering the mutual-inductance instability mode \citep{Srensen2011InstabilityWakes} more agressively. As a result, the mixing process in the wake and the resulting re-energization becomes more effective. Similar to DIC, the helix approach alters the strength of vortices that are shed from the blades due to the varying pitch angles, inserting higher levels of instability in the resulting vortex rings. Furthermore, these variations are distributed asymmetrically over the rotor. %The time-averaged flow fields do not provide a very clear picture of the described phenomena. In order to get a better perspective, we need to study the phase-locked data.

Figure~\ref{fig:Ta_mke_fields} offered a glance at the magnitude of velocity fluctuations  present in the wake for the different test cases. However, given the definition of the time-averaged Reynolds stress from \Cref{eq:reynolds_stress}, it does not distinguish between periodic and random contributions towards the entrainment of energy in the wake. \citet{Lignarolo2015Tip-vortexWakes} showed that the net periodic contribution (e.g., from tip vortices) to the transport of energy to the wake was nearly zero. Essentially, the tip vortices shield the wake and thereby prevent mixing of the wake with the surrounding flow. Following the completion of the leapfrogging motion, positive entrainment of energy was observed, which was solely dependant on random turbulent fluctuations in the flow. Using phase-averaged PIV measurements, we can study the contribution of turbulent fluctuations to the recovery of the wake in a similar manner for the helix cases.

\subsection{Phase-averaged flow analysis}
The PIV measurements were obtained in a time-resolved manner, and would therefore be suitable for dynamic analysis by themselves, were it not for the fact that the seeding of the flow tracers in the wake is not entirely uniform. As a result, less particles are recorded in some areas of the wake which means the measurement convergence might be lower in those parts. This problem can be overcome by phase-averaging the particle data during post-processing. This means that for each phase, more particles are available to each bin on the Cartesian grid, leading to a higher convergence of the velocity components.

\begin{figure}[t!]
    \hfill
    \includegraphics[scale=1.,valign=t]{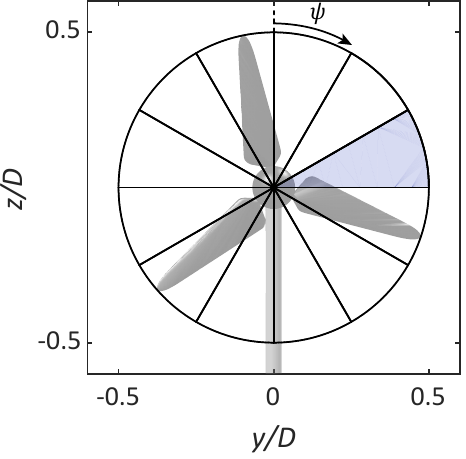}\hfill
    \includegraphics[scale = 1,valign=t]{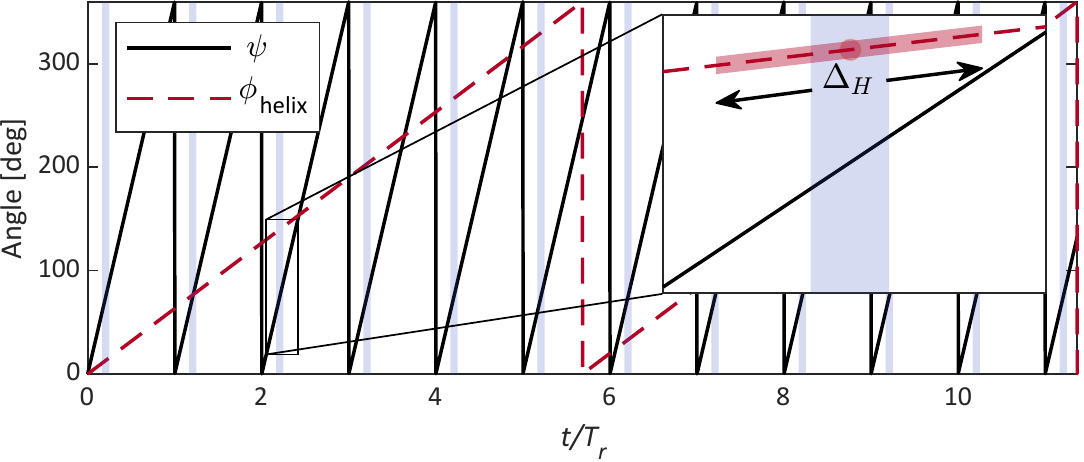}
    \hfill
    \caption{Graphical illustration of the phase-averaging procedure in case of a CCW helix implementation. Data is initially binned based on the measurement of the rotor azimuth position $\psi$, with bin sizes of $\SI{30}{\degree}$. Furthermore, measurements of the helix approach are binned based on a second criterion, which is the so called helix azimuth angle $\phi_\text{helix}$. For a particular rotor azimuth bin, indicated by the blue shaded area, the corresponding ranges of the helix azimuth bin are determined with bin width $\Delta_H$.}
    \label{fig:phase-locking procedure}
\end{figure}

For the baseline case, we phase-averaged the data using the measurement of the rotor azimuth angle $\psi$. The rotor was divided in twelve equal parts of $\SI{30}{\degree}$, as seen on the left side of \Cref{fig:phase-locking procedure}. Hence, the dynamics of the flow are now represented by twelve consecutive flow fields. On average, over 800 frames (i.e., sets of particle data at a specific time instant) were used for the averaging of each azimuth bin.

Phase-averaging measurement data for the helix approach is less straightforward than the previous case. Again, the data initially needs to be sorted using the rotor azimuth in order to retain the flow information from root and tip vortices, for example. However, due to the individual pitch action, this only delivers the average performance of the helix over a single rotation as shown in \citep{FlorianHeckmeier}. To visualize the dynamic effects of the helix approach, we used a second binning criterion that incorporates the angle between the yaw and tilt moments from \Cref{fig:MBCtransform}. This angle, referred to as the helix azimuth angle, is given by
\begin{equation}
    \phi_\text{helix} = \tan^{-1}\left(\frac{\theta_\text{tilt}}{\theta_\text{yaw}}\right).
\end{equation}

The new binning procedure is visualized on the right hand side of \Cref{fig:phase-locking procedure}, which shows the two sawtooth waves for the rotor azimuth $\psi$ and helix azimuth $\phi_\text{helix}$. Unfortunately, the frequencies of the rotor and the two helix test cases do not match up perfectly. For each CCW cycle of the helix, the rotor completes $f_r/f_e=5.68$ revolutions. This means that we have to go through three helix cycles, before both the rotor and helix azimuth angles return to the initial conditions. Consequently, the number of bins used for phase-averaging is expressed by (\textit{\# of helix cycles}) x (\textit{\# of azimuth bins}). For the CW case, this resulted in 168 bins, while 204 bins were obtained for the CCW case due to the slightly lower excitation frequency.

A single rotor azimuth bin is plotted in \Cref{fig:phase-locking procedure} for each consecutive revolution of the rotor by the light blue shaded area. Zooming in on one these azimuth bins, we see that the range of $\phi_\text{helix}$ corresponding to this bin is limited. To increase the number of samples for each bin, an additional variable is introduced representing the width of the helix azimuth bin $\Delta_H$. Investigating the effect of this variable on the resulting velocity fields, a bin width of $\Delta_H=\SI{40}{\degree}$ was seen to provide converged results. Summarizing the binning procedure: for each consecutive rotor azimuth bin, the corresponding helix azimuth is determined, after which all frames that belong to this rotor azimuth and fall within the bin of the helix azimuth are collected for the averaging procedure. On average, around 150 frames were used to obtain the velocity fields of each bin. 

\begin{figure}[hbt!]
    \begin{subfigure}{1\textwidth}
        \centering
        \caption{}
        \includegraphics[scale=1]{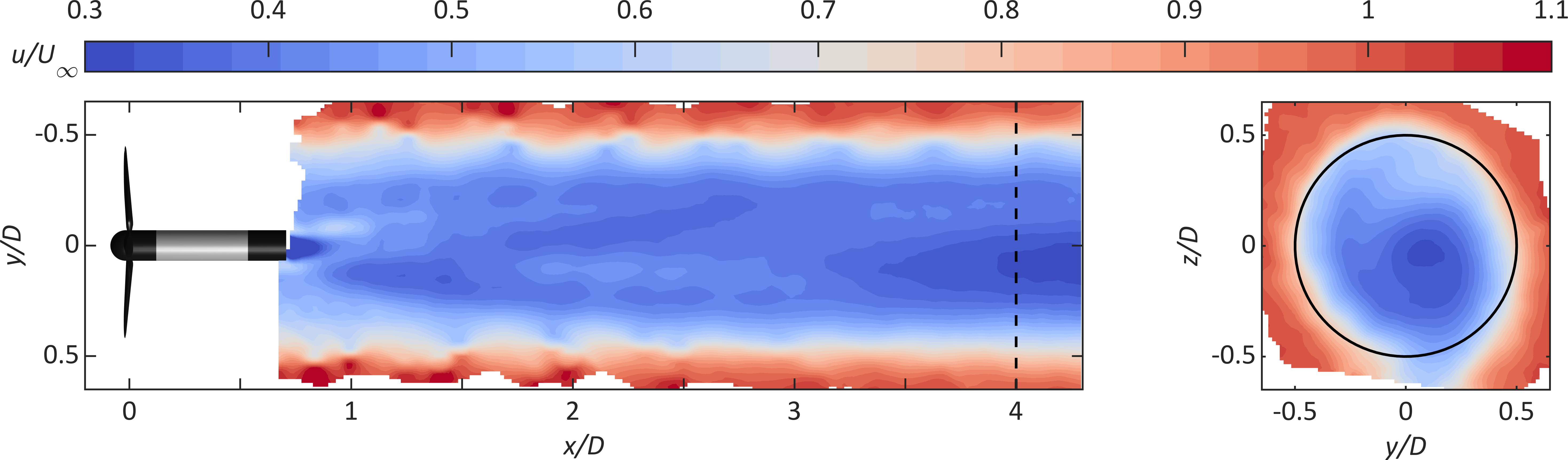}
    \end{subfigure}
    \begin{subfigure}{1\textwidth}
        \centering
        \caption{}
        \includegraphics[trim = {0, 0.7cm, 0, 0.92cm}, clip,scale=1]{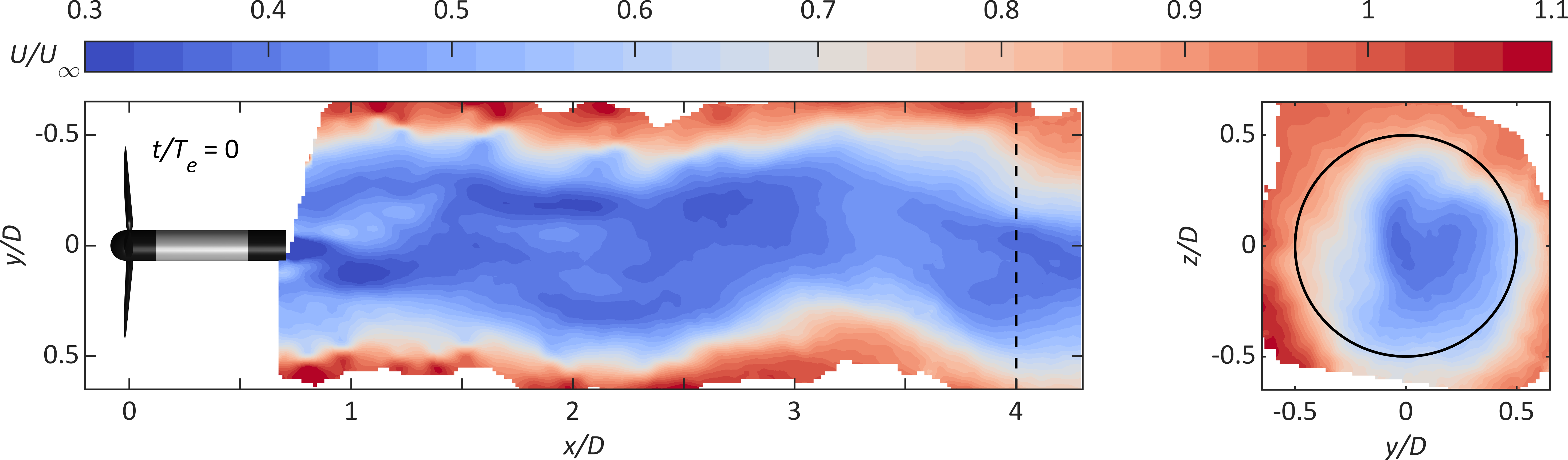}
        \includegraphics[trim = {0, 0.7cm, 0, 0.92cm}, clip,scale=1]{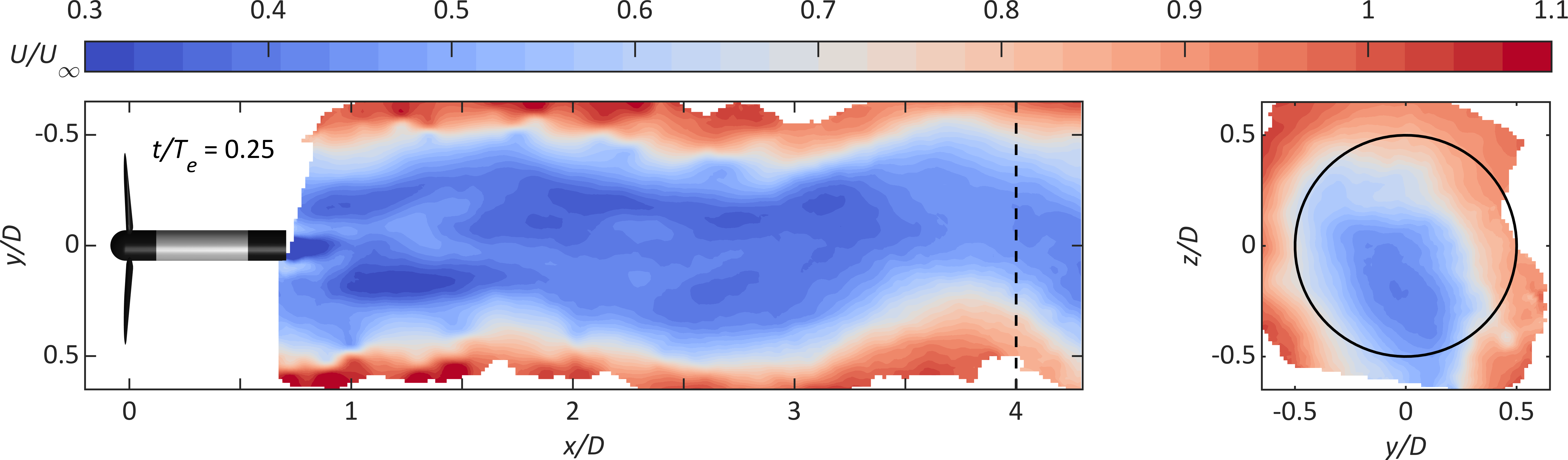}
        \includegraphics[trim = {0, 0.7cm, 0, 0.92cm}, clip,scale=1]{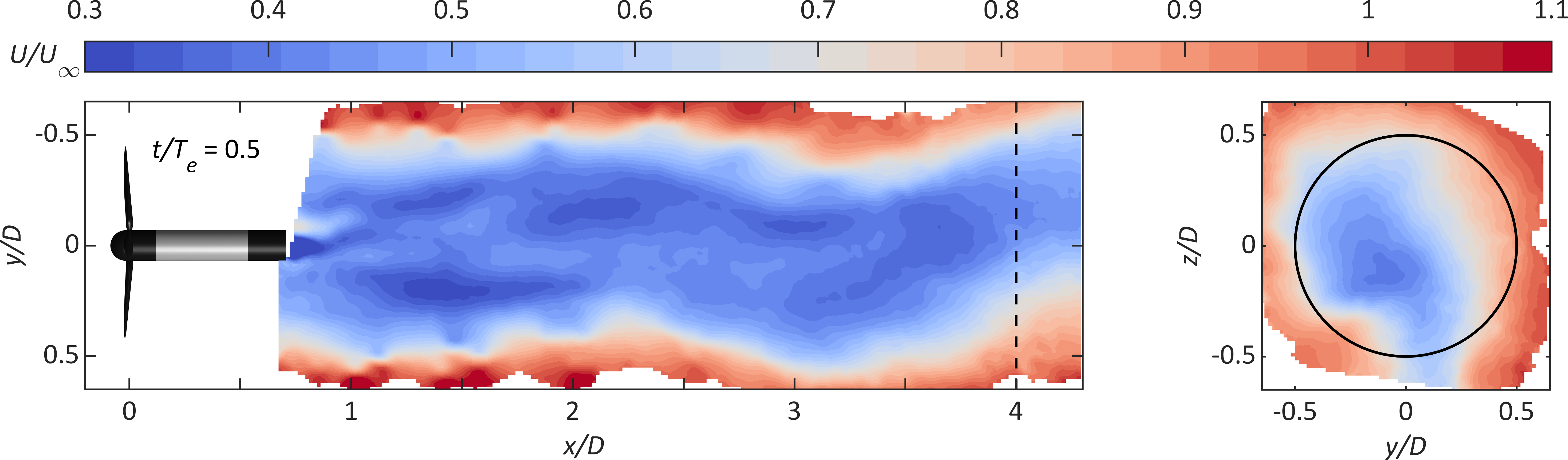}
        \includegraphics[trim = {0, 0cm, 0, 0.92cm}, clip,scale=1]{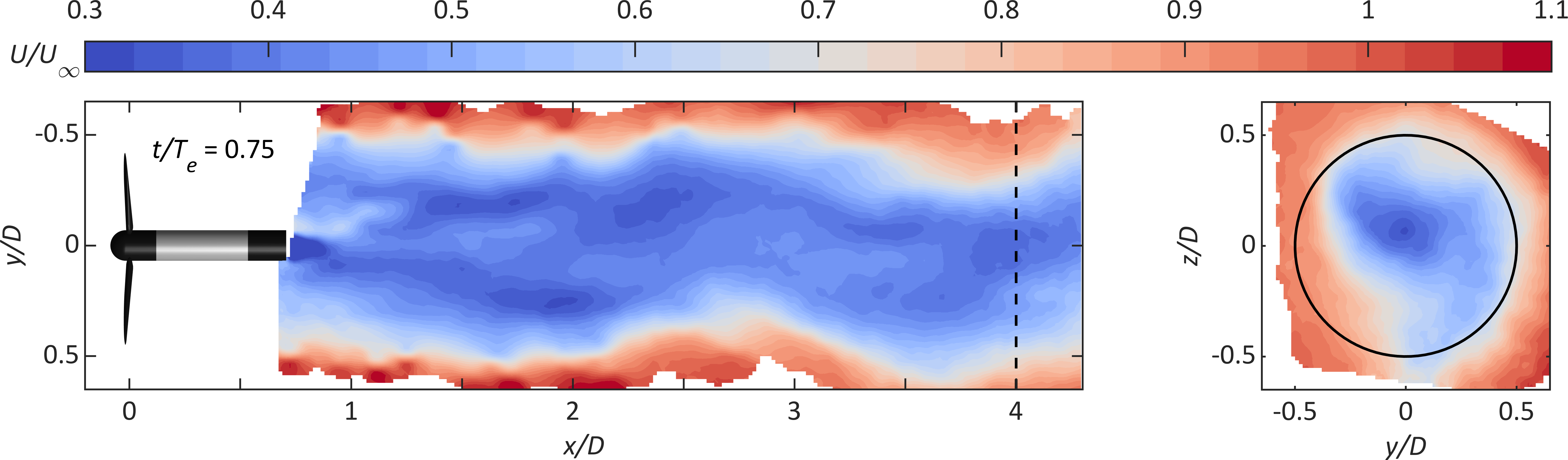}
    \end{subfigure}
    \caption{Phase-averaged velocity contours $u/U_\infty$ at turbine hub height ($z/D=0$) and of a cross-section of the wake at $x/D=4$. A single frame of baseline operation (a) is compared to multiple phases of a single CW helix cycle (b).}
    \label{fig:velocity_pa}
\end{figure}

The phase-averaged velocity fields of the baseline and CW helix cases are compared in \Cref{fig:velocity_pa}. The wake of the turbine operating under greedy control is similar to its time-averaged counterpart, apart from additional details in the shear layer of the wake. Here, we can already observe the presence of the tip vortices. The velocity fields of the CW helix are given at four different phases of the helix cycle. The top view provides a clear picture of the forced meandering caused by the varying yaw and tilt moments. The CW movement of the wake is also clearly seen through the cross-sections at $x/D=4$.

\begin{SCfigure}
    \centering
    \includegraphics[scale=1]{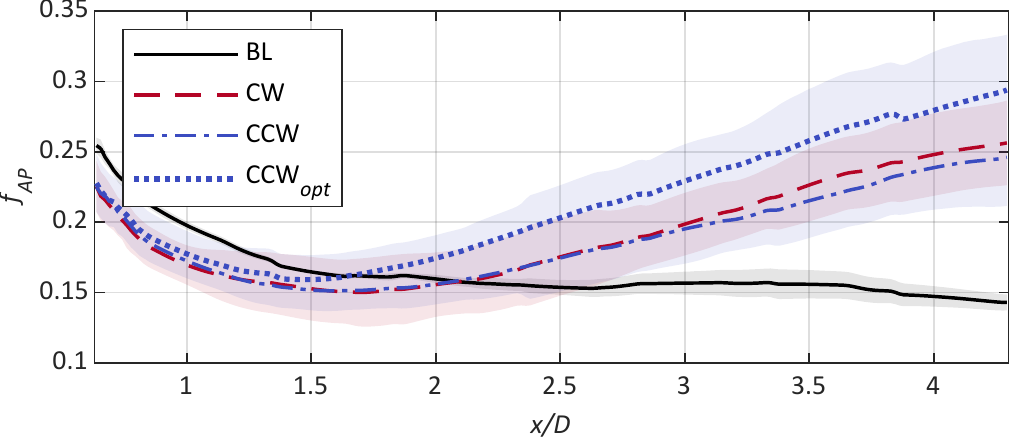}
    \caption{Evolution of the fraction of available power over the downstream distance $x/D$. The $f_\text{AP}$ is based on the power extracted by a hypothetical turbine rotor using the phase-averaged velocity fields. The shaded areas indicate the variation in power levels over the different phase bins, expressed by $\pm2\sigma$.}
    \label{fig:rel_power_wake}
\end{SCfigure}

The performance of the helix approach compared to baseline operation can already be quantified using the streamwise velocity fields as shown in \Cref{fig:velocity_pa}. The fraction of available power $f_\text{AP}$ provides an indication of the amount of power that is available compared to the free stream inflow to a hypothetical turbine in the wake \citep{Vollmer2016EstimatingStudy}. This performance criterion is given by
\begin{equation}
    \centering
    f_{\mathrm{AP}}(x,r,\psi) = \int_0^{2\mathrm{\pi}}\int_0^R u^3(x,r,\psi)/U_{\mathrm{\infty}}^3 r \mathrm{d}r\mathrm{d}\psi,
\end{equation}
with $x$ denoting the distance downstream, $r$ the radial distance measured from the rotor center, and azimuth angle $\psi$. Figure~\ref{fig:rel_power_wake} presents the $f_\text{AP}$ for each of the different test cases from \Cref{tab:test_cases}. In the figure, the CCW helix implementation with an average amplitude of $A_\theta=\SI{2.4}{\degree}$ is referred to as $\text{CCW}_\text{opt}$. All three helix implementations initially see similar levels of available power in the wake, though below baseline operation. Under baseline operation the power decreases at a slower rate until it settles to a near constant value around $x/D=2$, showing no signs of recovery in the wake domain that was measured. The CW and CCW implementations reach their minimum level between $x/D=1.5$ and $x/D=2$, after which both start increasing at a similar rate. The variation in power level depending on the different phases in the helix actuation sequence due to the horizontal and vertical movement of the wake is also visualized by the shaded areas. The CW method was able to reach a slightly higher power level, likely as a result of the higher pitch amplitude. If we also compare it to the available power with $\text{CCW}_\text{opt}$, we see an even earlier recovery of the wake ($x/D=1.5$) and higher power levels further downstream compared to the two other helix cases.

\subsection{Kinetic energy entrainment}
Next, we consider the phase-averaged entrainment into the wake due to shear stresses caused by random fluctuations. This process has been identified as the dominant factor for re-energizing the wake, following the tip vortex breakdown \citep{Lignarolo2015Tip-vortexWakes}.
The phase-averaged Reynolds stress in the vertical plane for a single phase $\phi$ is defined as 
\begin{equation}
\label{eq:pa_reynolds_stress}
    \langle u_s w_s\rangle_\phi = \frac{\sum_{k=1}^N\left(u(t_{k,\phi})-\langle u\rangle_\phi\right)\left(w(t_{k,\phi})-\langle w\rangle_\phi\right)}{N},
\end{equation}
with $\langle u\rangle_\phi$ and $\langle w\rangle_\phi$ the phase-averaged in-plane velocity components, and $t_{k,\phi}$ the sampling time for a velocity component in phase bin $\phi$. In this case, the periodic fluctuations are now an inherent part of the phase-averaged velocity $\langle u\rangle_\phi$. However, the phase-averaging procedure adopted in this paper did not result in strictly phase-locked measurements, as the phase bins contain a collection of smaller phases.   

\begin{figure}[hbt!]
    \begin{subfigure}{1\textwidth}
        \caption{}
        \includegraphics[scale=1]{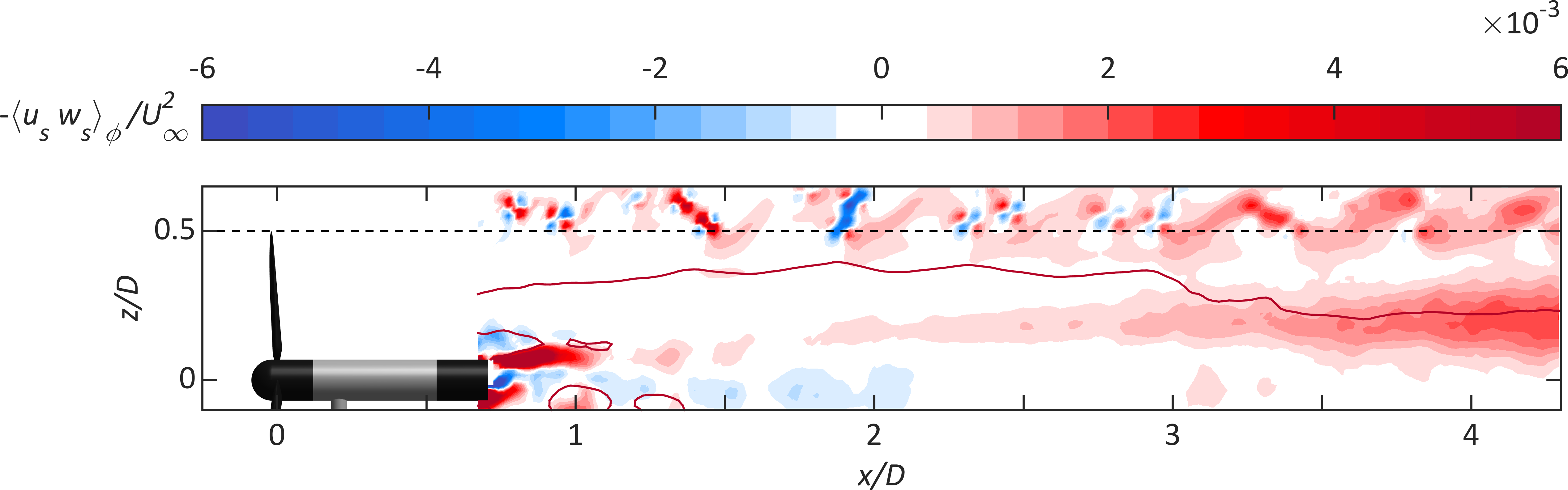}
    \end{subfigure}
    \begin{subfigure}{1\textwidth}
        \caption{}
        \includegraphics[trim = {0, 0.65cm, 0, 1.5cm}, clip,scale=1,valign=t]{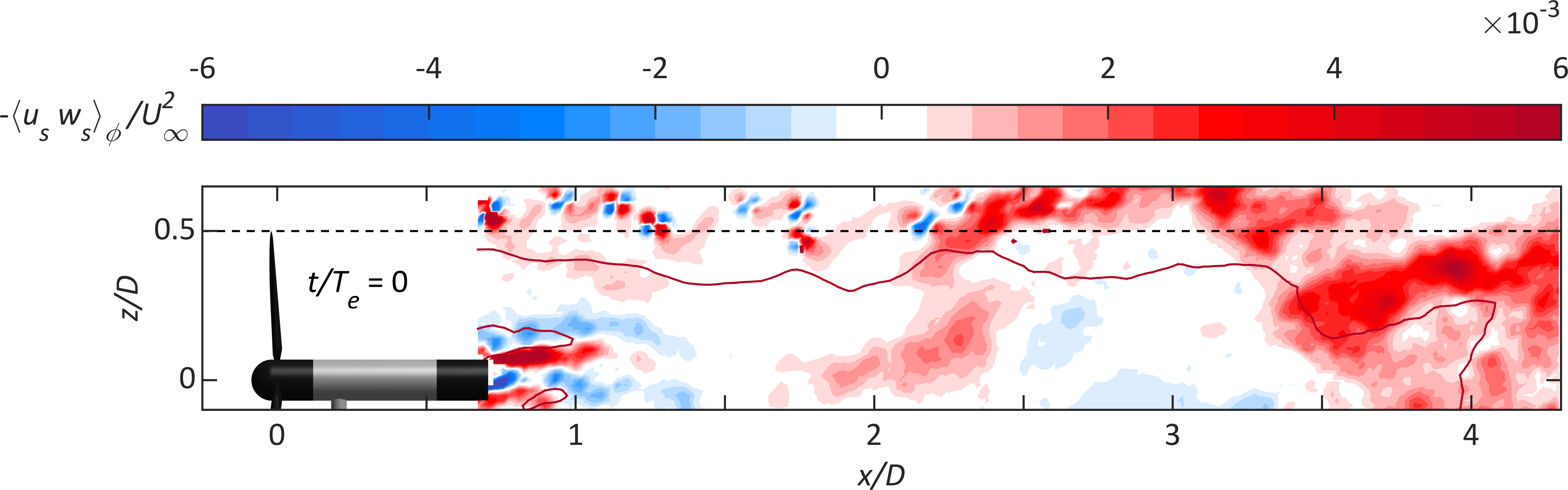}
        \hspace{0.5cm}
        \includegraphics[trim = {0, 0.75cm, 0, 0}, clip,scale=1,valign=t]{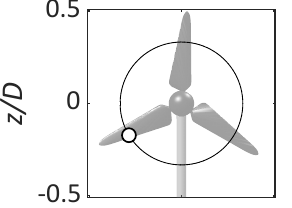}\\
        \includegraphics[trim = {0, 0.65cm, 0, 1.5cm}, clip,scale=1,valign=t]{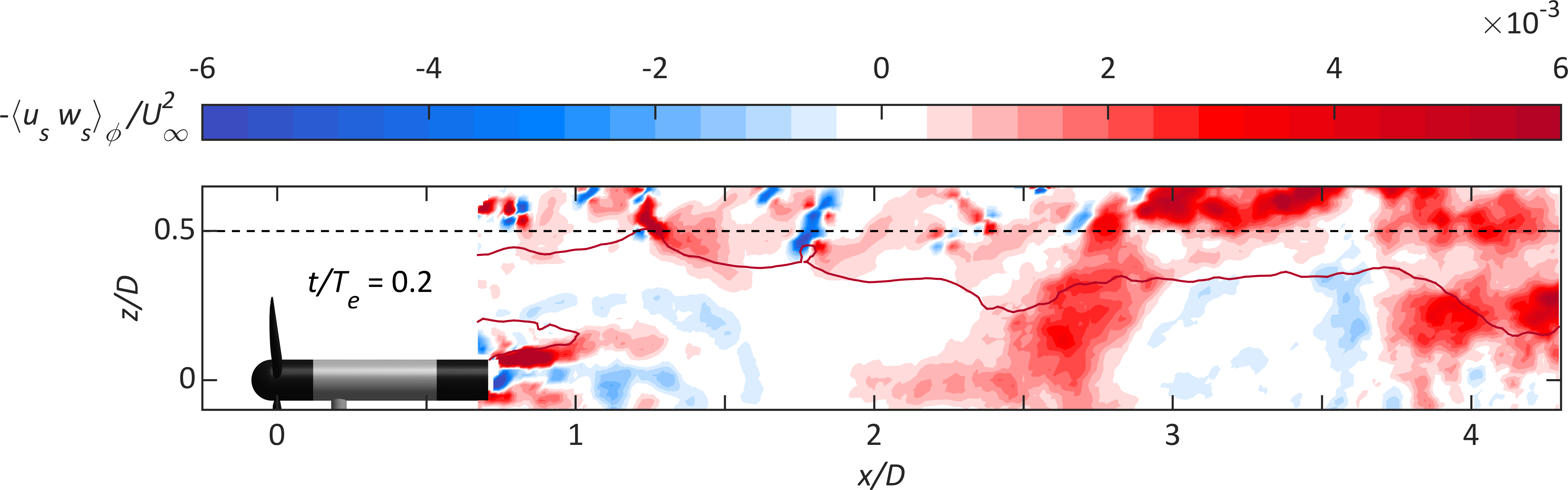}
        \hspace{0.5cm}
        \includegraphics[trim = {0, 0.75cm, 0, 0}, clip,scale=1,valign=t]{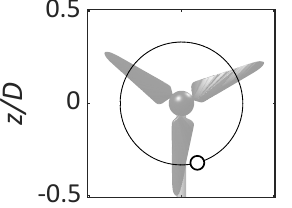}\\
        \includegraphics[trim = {0, 0.65cm, 0, 1.5cm}, clip,scale=1,valign=t]{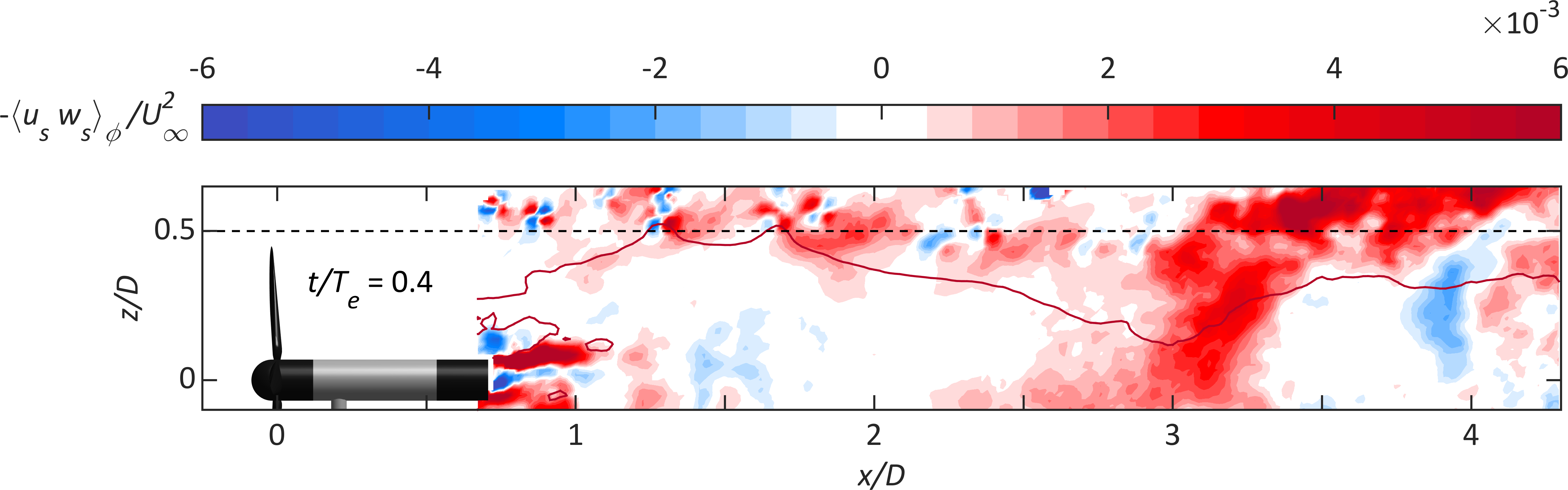}
        \hspace{0.5cm}
        \includegraphics[trim = {0, 0.75cm, 0, 0}, clip,scale=1,valign=t]{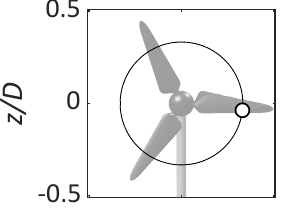}\\
        \includegraphics[trim = {0, 0.65cm, 0, 1.5cm}, clip,scale=1,valign=t]{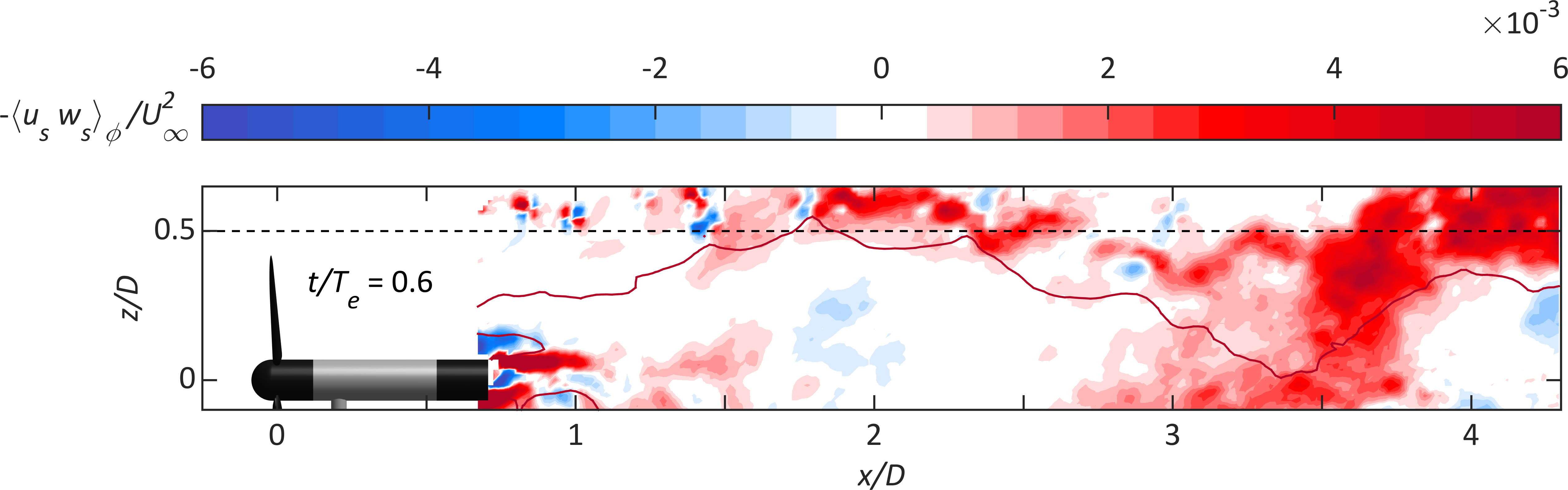}
        \hspace{0.5cm}
        \includegraphics[trim = {0, 0.75cm, 0, 0}, clip,scale=1,valign=t]{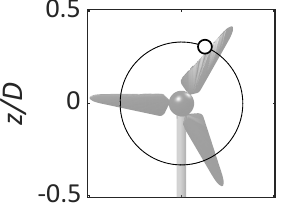}\\
        \includegraphics[trim = {0, 0cm, 0, 1.5cm}, clip,scale=1,valign=t]{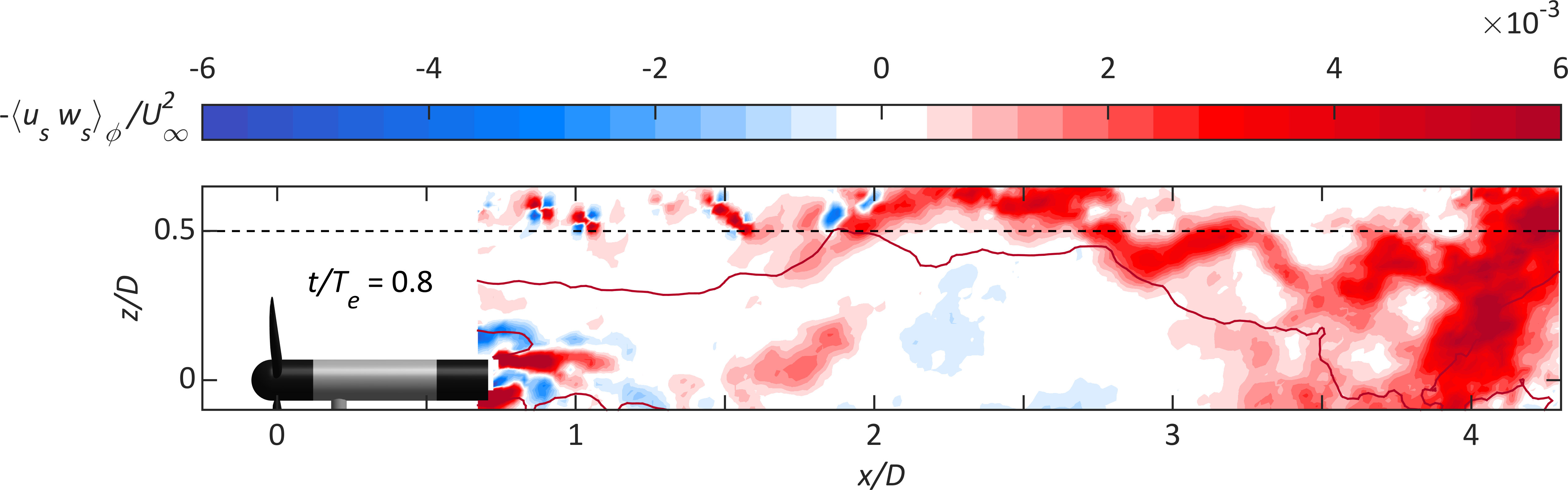}
        \hspace{0.5cm}
        \includegraphics[scale=1,valign=t]{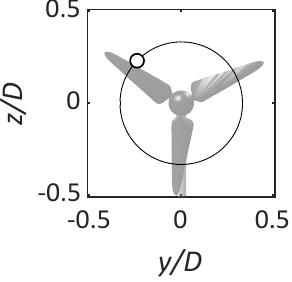}
    \end{subfigure}
    \caption{Phase-averaged Reynolds stress in the vertical plane ($y/D=0$) for baseline operation (a) and the CCW helix approach (b). A negative definition of the Reynolds stress was used, such that positive values indicate the entrainment towards the negative $z$ direction (i.e., downward). We are mostly interested in the energy that crosses the border of the wake shown by the dashed black line. The solid black contour line indicates a streamwise velocity of $\langle u\rangle_\phi/U_\infty=0.5$. The different phases of the helix approach are illustrated on the right hand side by the location of the resultant tilt and yaw moment over the rotor plane.}
    \label{fig:entrainment_pa}
\end{figure}

Multiple phases of the phase-averaged shear stress are visualized in \Cref{fig:entrainment_pa}. The dashed line indicates the edge of the wake, meaning that any red areas crossing this line represent positive flux of kinetic energy into the wake. The solid black line in the figure indicates the streamwise velocity contour of $\langle u\rangle/U_\infty=0.5$. The flow from the baseline case shows how saddle points mark the location of subsequent tip vortices. Furthermore, the initial onset of two pairing vortices can be observed just before $x/D=1.5$. Judging from the sign flip of the shear stress a little further downstream, the leapfrogging motion has just been completed for another pair of vortices at that location. Moving further downstream, we now see a more positive contributions of the shear stress crossing over into the turbine wake. The second part of \Cref{fig:entrainment_pa} shows the phase-averaged Reynolds stress at different phases of the CCW helix cycle, which are marked by the location of the resultant tilt and yaw angles. Considering the first phase that is presented, we initially also see some saddle points indicating the presence of tip vortices. The shear stress quickly becomes more dominant compared to the baseline case as we move further downstream, concentrating in a larger area around $x/D=2.5$. Moving on to the next frames in the helix cycle, we can observe how more energy is accumulated as the flow moves downstream, and subsequently passes the wake boundary. This area of concentrated shear stress is seen to coincide with a downward movement of the wake, which is marked by the velocity contour. 

\begin{figure}[t!]
    \centering
    \includegraphics[trim = {0, 0.6cm, 0, 0}, clip,scale=1]{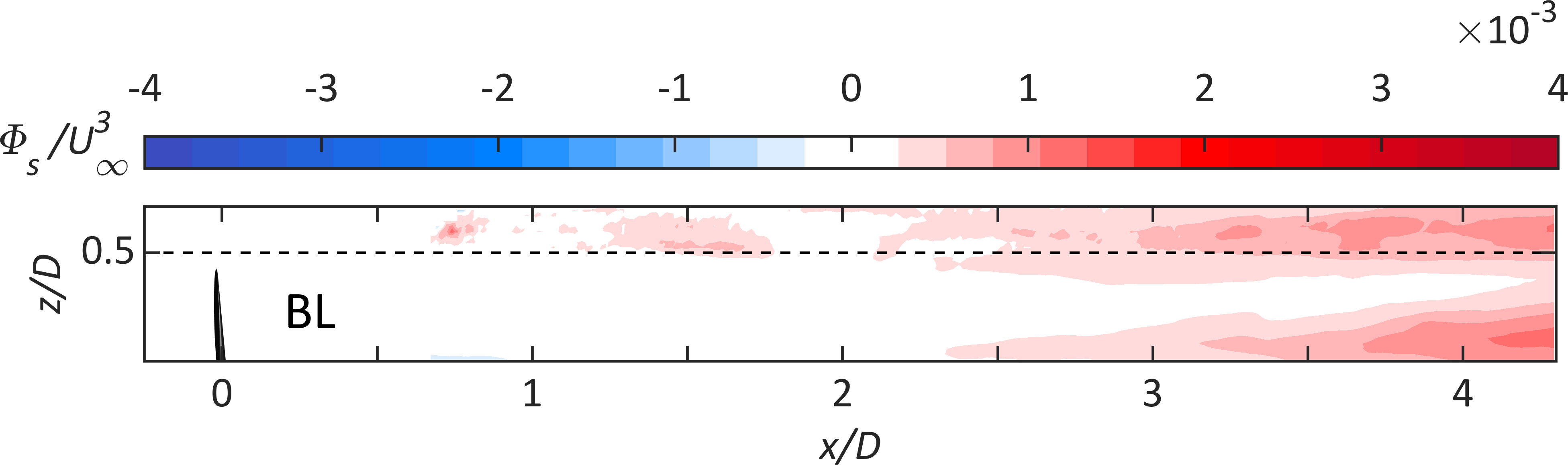}\\
    \includegraphics[trim = {0.54cm, 0.6cm, 0, 1.2cm}, clip,scale=1]{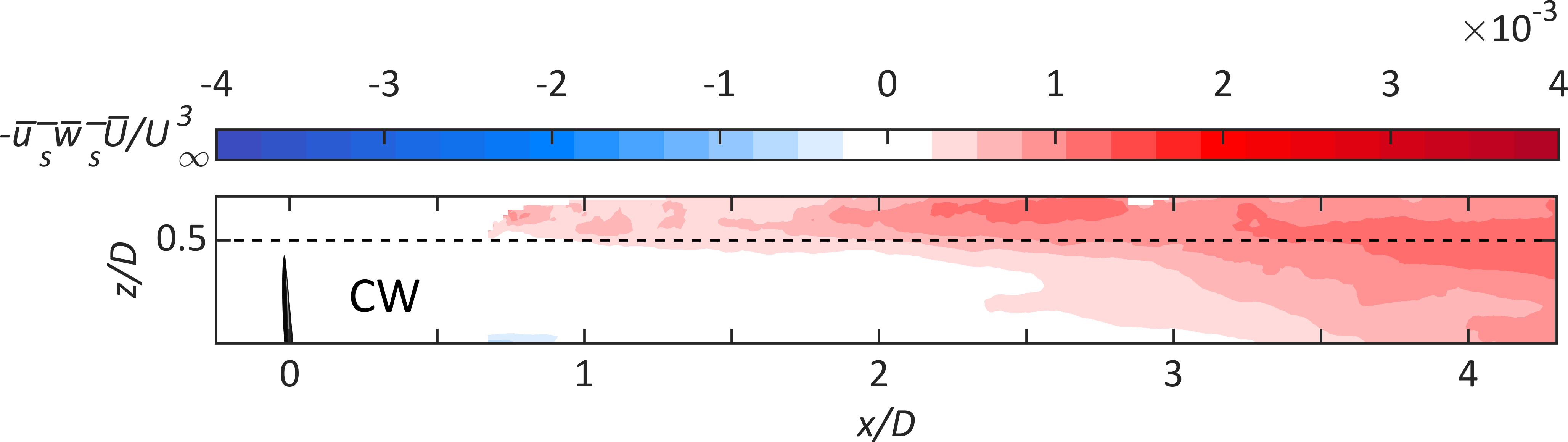}\\
    \includegraphics[trim = {0.54cm, 0cm, 0, 1.2cm}, clip,scale=1]{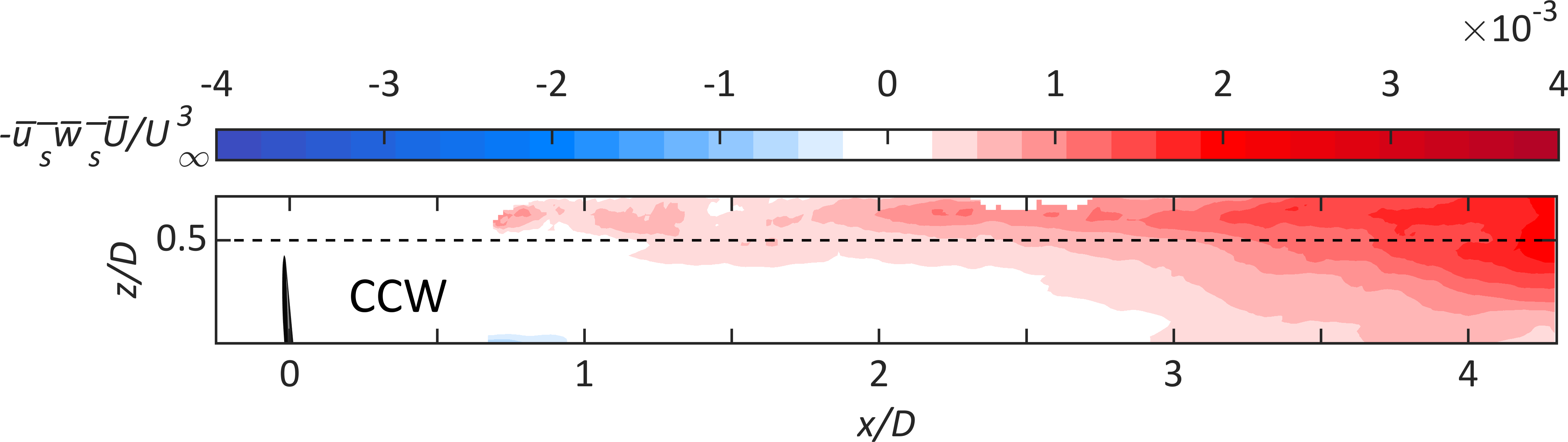}
    \caption{Vertical flow field ($y/D=0$) of the normalized flux of mean kinetic energy due to random fluctuations into the wake region. A negative definition of the flux was taken, meaning that red areas indicate flux of kinetic energy in the negative $z$ direction (downward). We are primarily interested in the flux crossing the boundary of the wake, marked by the dashed line. The figure shows a comparison between baseline operation, CW helix, and CCW helix.}
    \label{fig:entrainment_random}
\end{figure}

Figure~\ref{fig:entrainment_pa} illustrated how the entrainment of kinetic energy as shown by the phase-averaged Reynolds stress varied over different phases of the helix cycle. By averaging these results over all considered phases, we can obtain an expression for the net entrainment of kinetic energy into the wake as a result of random turbulent motions in the flow. The average Reynolds stress due to random fluctuations over all phases is given by 
\begin{equation}
    \overline{u_s w_s} = \sum_{\phi=1}^{N_f}\frac{\langle u_s w_s\rangle_\phi}{N_f},
\end{equation}
for a number of phases $N_f$. Multiplying this term with the time-averaged velocity provides the flux of mean kinetic energy due to random fluctuations:
\begin{equation}
    \Phi_s = -\overline{u_s w_s}\bar{u}.
\end{equation}
Figure~\ref{fig:entrainment_random} presents the mean transport of kinetic energy into the wake as a result of random fluctuations for the three considered test cases. Comparing the baseline result to the one from \Cref{fig:Ta_mke_fields}, we no longer observe the initial positive flux followed by the sign flip, which indicates the leapfrogging location. This initial area, which was dominated by periodic fluctuations resulting from the tip vortices, does not have a positive net contribution to the recovery of the wake \citep{Lignarolo2015Tip-vortexWakes}. This is confirmed by \Cref{fig:entrainment_random}, which shows that the entrainment is dominated by the region after the leapfrogging occurrence. Both implementations of the helix approach show a clear increase in the kinetic energy flux. The helix is able to achieve higher levels of mean kinetic energy flux due to the meandering movement of the wake. Furthermore, the entrainment starts at a shorter distance from the rotor plane. We believe this to be the result of enhanced interaction between tip vortices, which will be considered in the next section.  

\subsection{Tip vortex evolution}
The effect of the tip vortex instability on the recovery of the wake has been mentioned several times before in this paper. Using simulations with a vortex-particle-mesh code, \citet{CoqueletThesis} showed how the meandering of the wake was triggered differently depending on the resolution of the simulations. At low to intermediate resolution, the wake was destabilized by Kelvin-Helmholtz instabilities, resulting in a meandering wake. When a sufficiently high resolution was used to fully resolve the tip vortices, the manner of destabilization changed to vortex pairing induced instabilities. The phase-averaged PIV flow fields also provide a good opportunity to study how the helix implementation affects the tip vortex pairing and initializes the destabilization of the wake. 

\begin{figure}[hbt!]
    \centering   
    \includegraphics[trim = {0, 0.65cm, 0, 0}, clip,scale=1,angle=90]{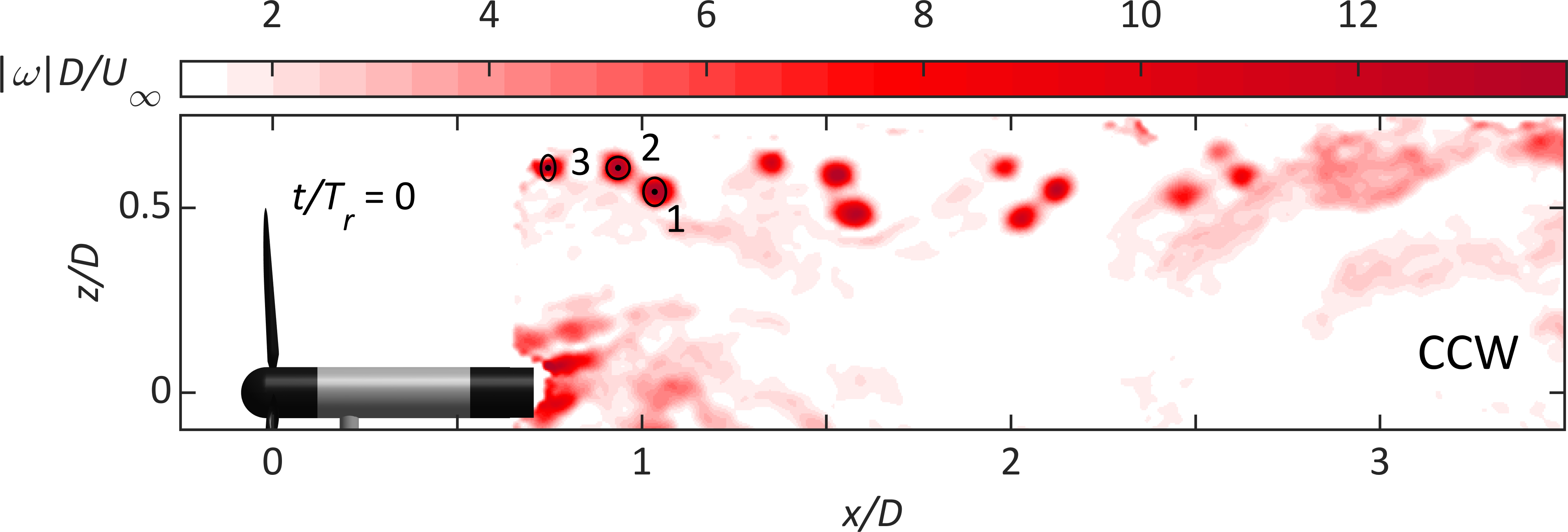}
    \includegraphics[trim = {0, 0.65cm, 0, 0.72cm}, clip,scale=1,angle=90]{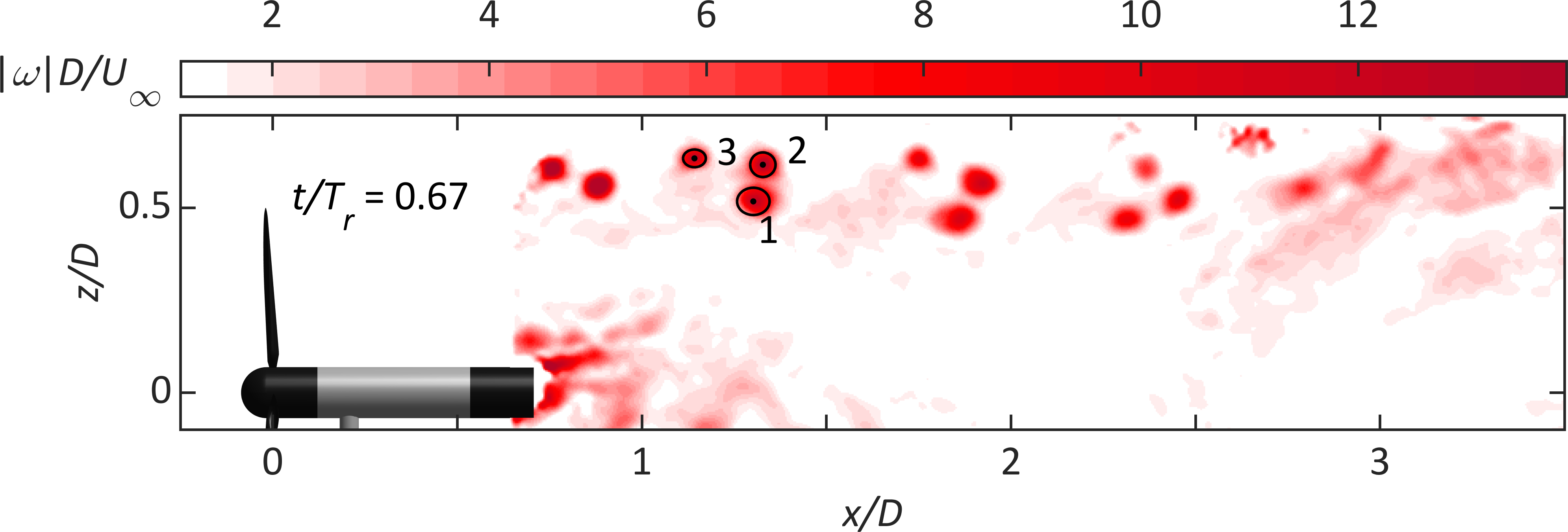}
    \includegraphics[trim = {0, 0.65cm, 0, 0.72cm}, clip,scale=1,angle=90]{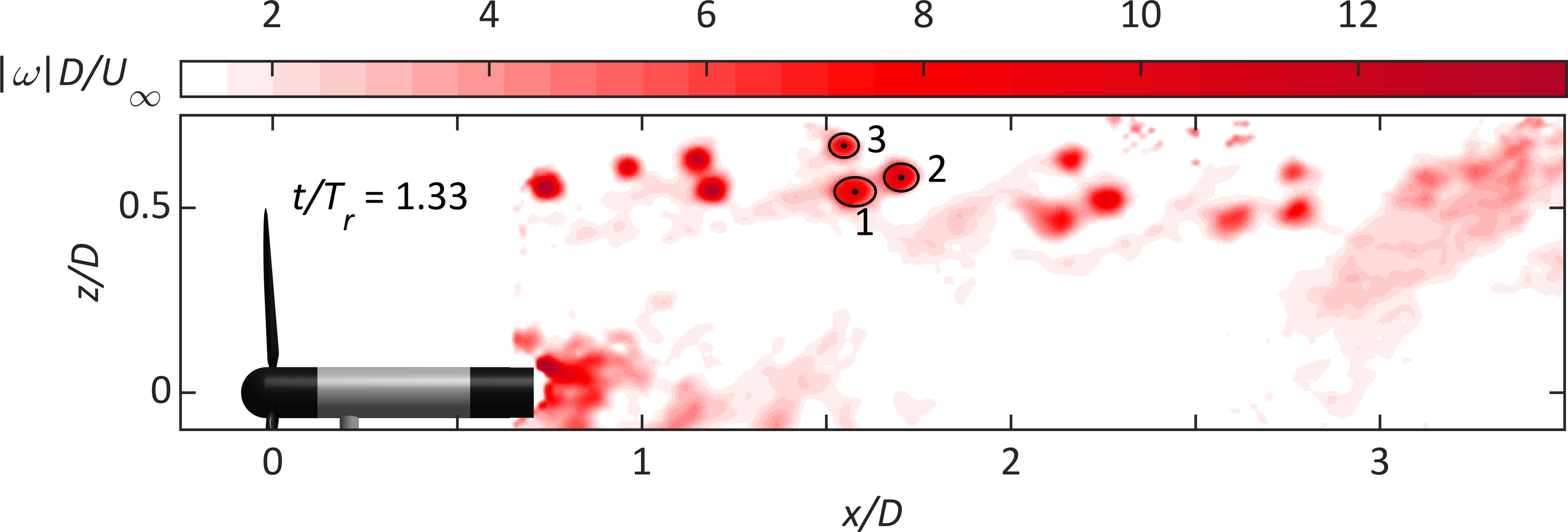}
    \includegraphics[trim = {0, 0.65cm, 0, 0.72cm}, clip,scale=1,angle=90]{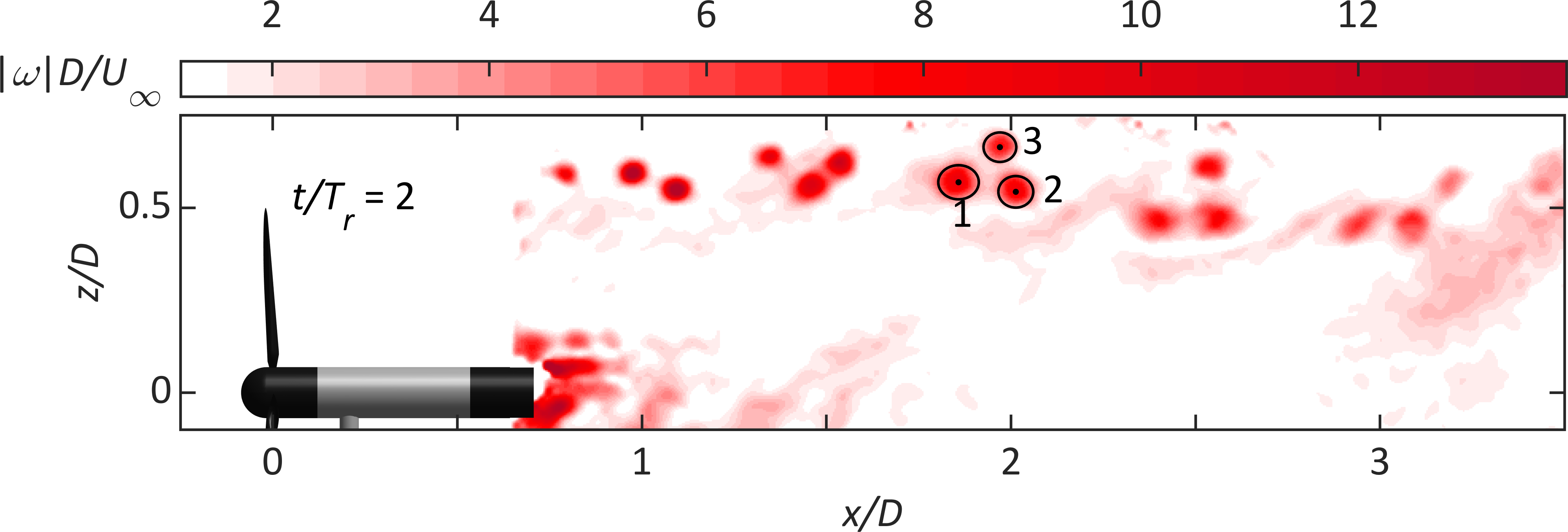}
    \includegraphics[trim = {0, 0cm, 0, 0.72cm}, clip,scale=1,angle=90]{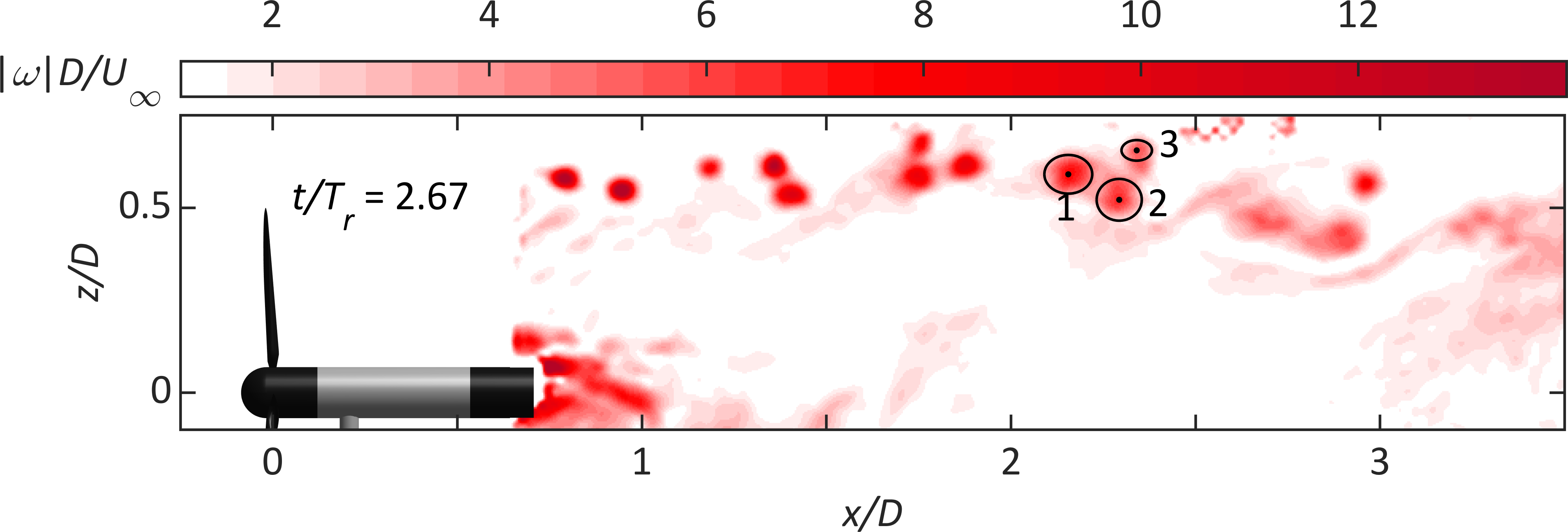}\\
    \vspace{0.5cm}
    \includegraphics[trim = {0, 0.65cm, 0, 0}, clip,scale=1,angle=90]{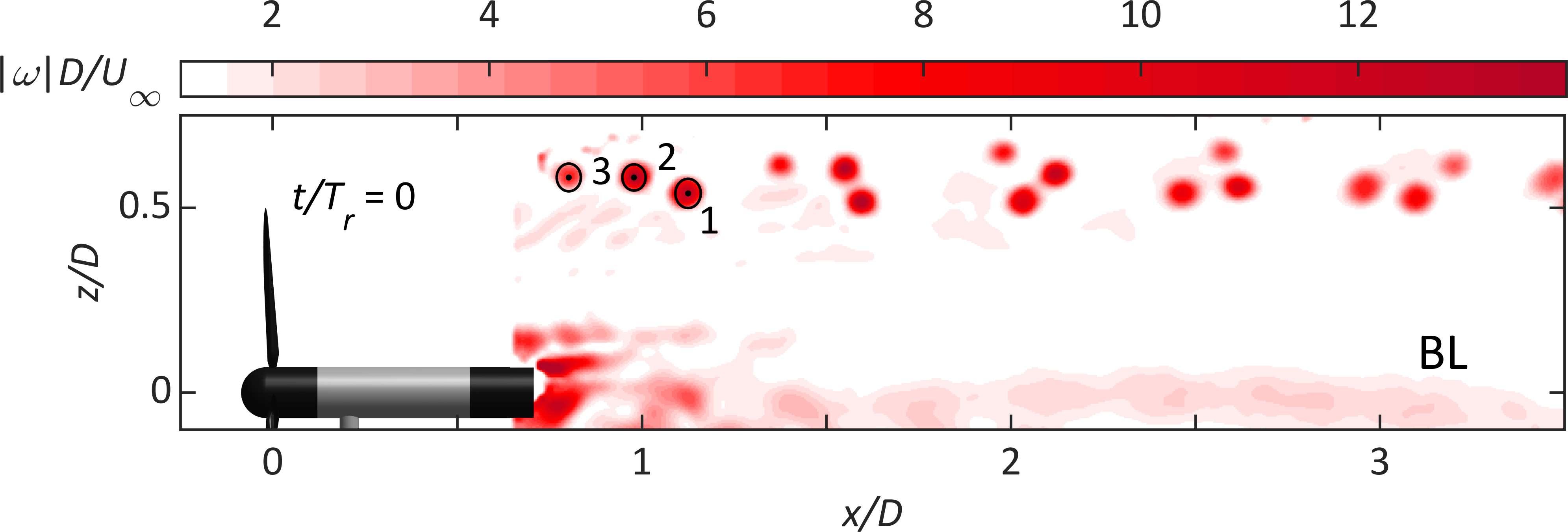}
    \includegraphics[trim = {0, 0.65cm, 0, 0.72cm}, clip,scale=1,angle=90]{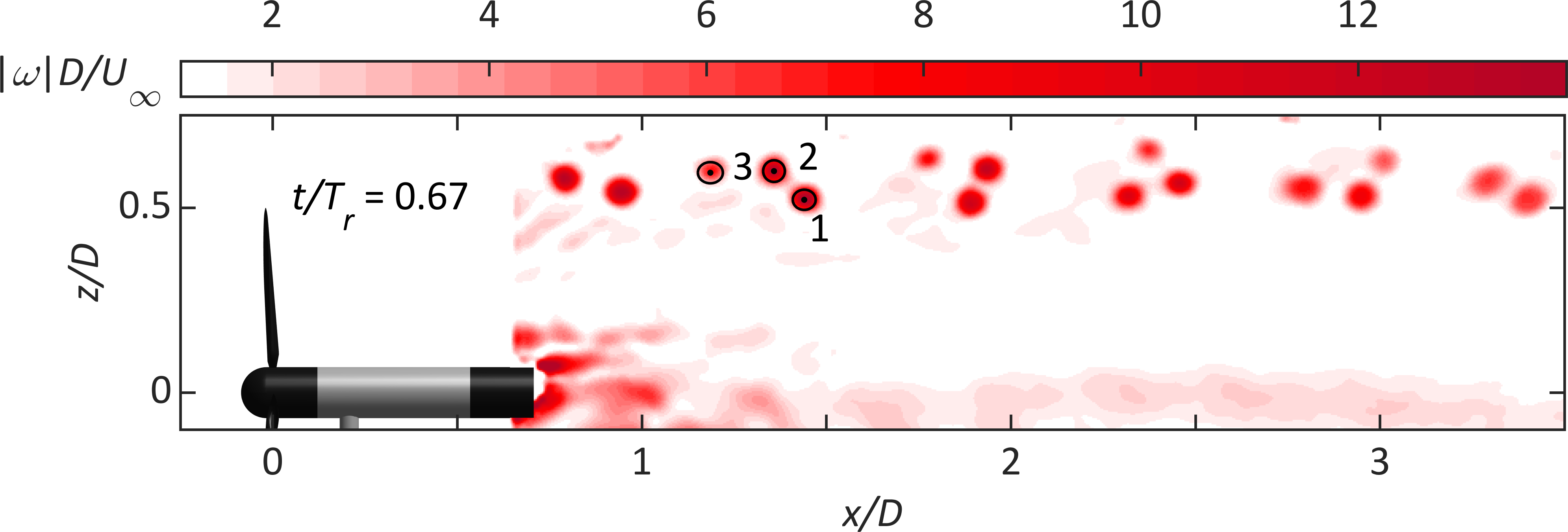}
    \includegraphics[trim = {0, 0.65cm, 0, 0.72cm}, clip,scale=1,angle=90]{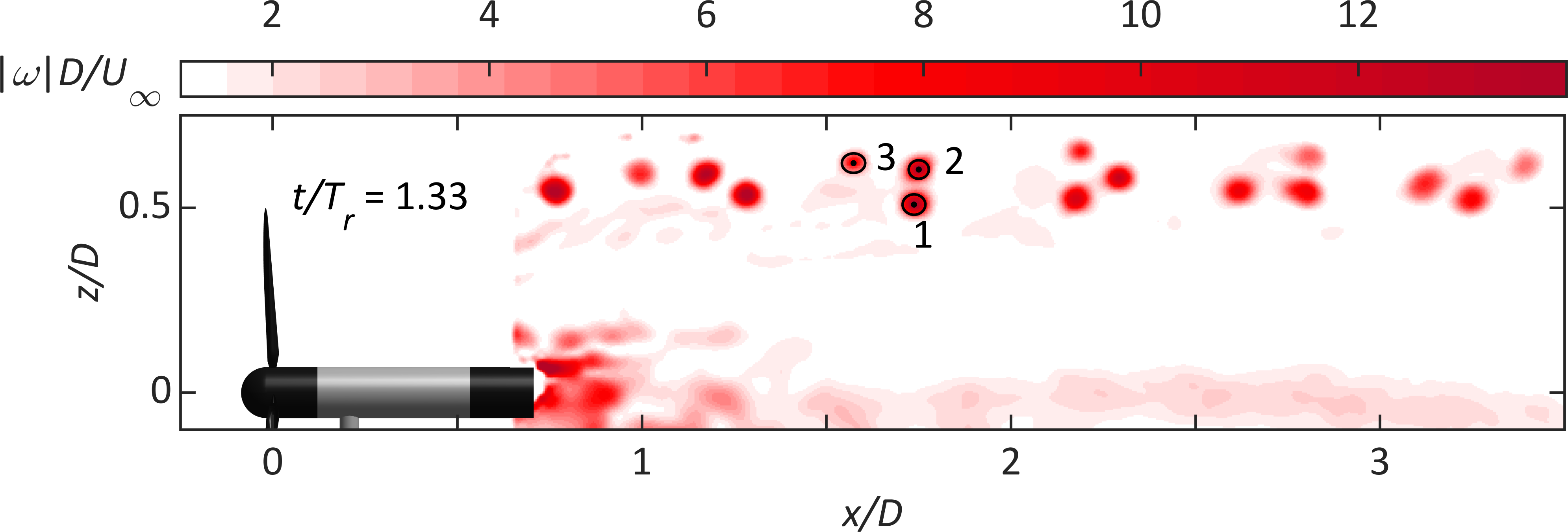}
    \includegraphics[trim = {0, 0.65cm, 0, 0.72cm}, clip,scale=1,angle=90]{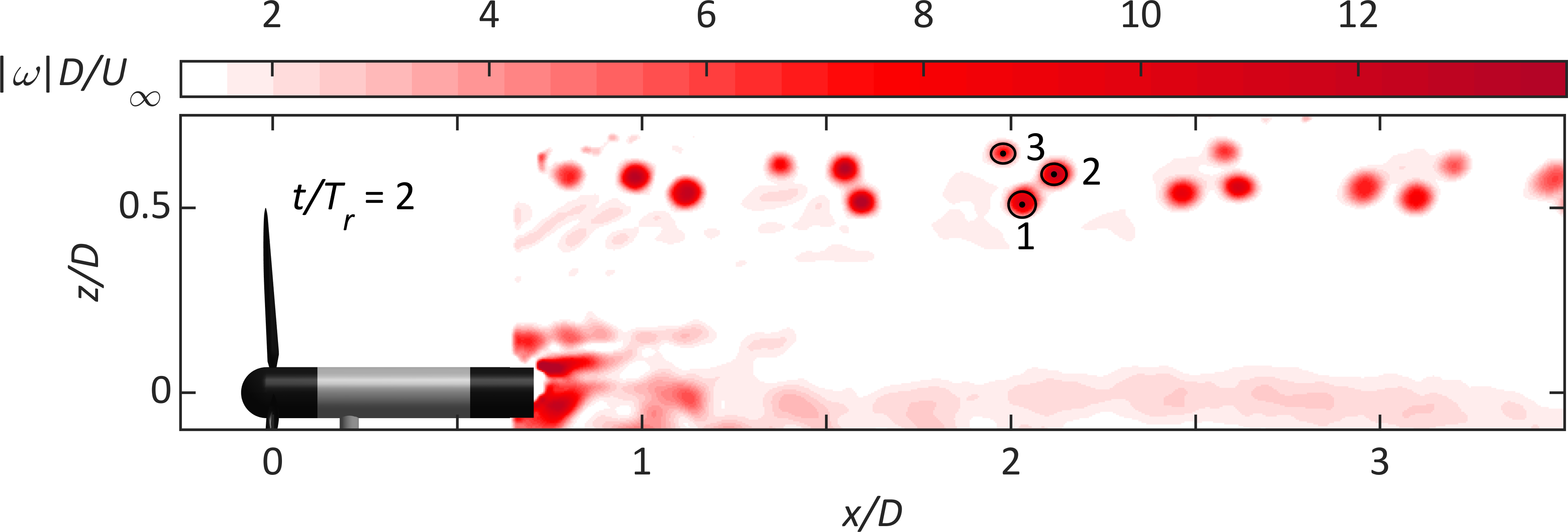}
    \includegraphics[trim = {0, 0cm, 0, 0.72cm}, clip,scale=1,angle=90]{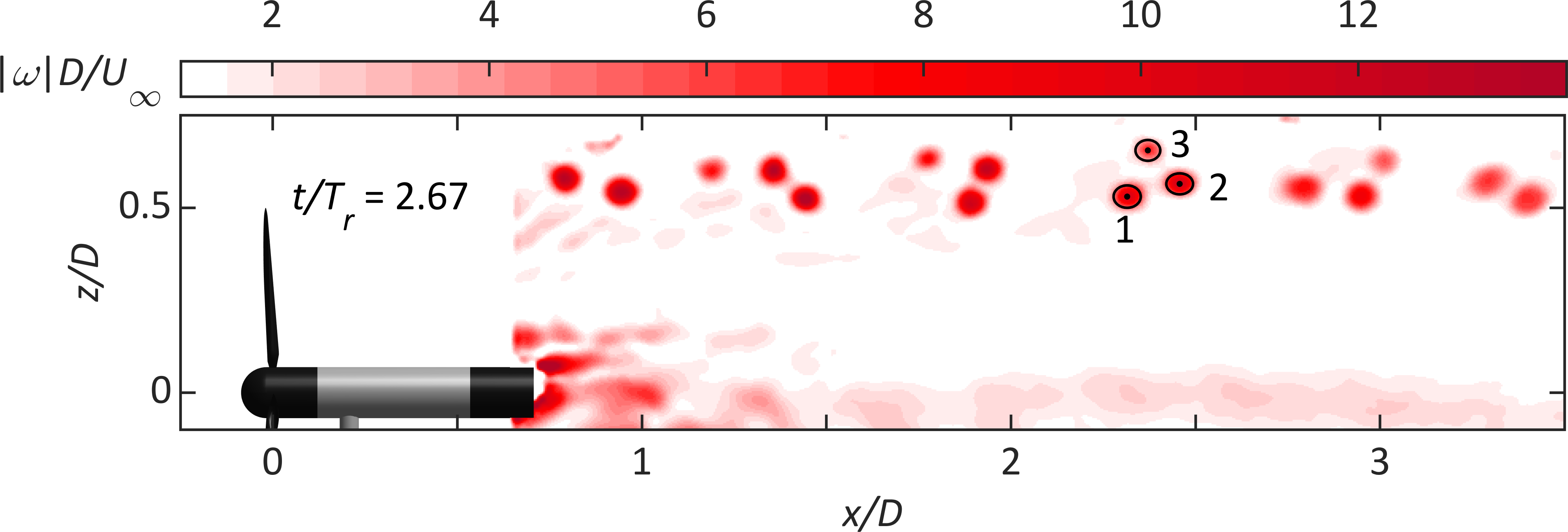}
    \caption{Evolution of the tip vortices, expressed by the normalized vorticity magnitude $|\omega|D/U_{\infty}$, as they travel downstream. The figure compares the interaction between the tip vortices for several phases, in case of baseline operation and the CCW helix approach. Increased interaction is visible for the helix case, resulting in accelerated diffusion of the tip vortices.}
    \label{fig:vortex_tracking}
\end{figure}

The evolution of the tip vortices is compared for the baseline and CCW helix case in \Cref{fig:vortex_tracking} using the three dimensional vorticity magnitude $|\omega|=\sqrt{\omega_x^2+\omega_y^2+\omega_z^2}$. For each test case, the tip vortices were tracked over time using a least-squares estimate of the displacement $\delta$ in $x$ and $y$ direction, based on two consecutive vorticity flow fields. This assumes that the displacement of vortices is mostly linear, not taking into account any rotation. To account for rotation, the least-squares estimates were subsequently corrected with a local peak finding algorithm. The peak location of the vorticity magnitude was taken as the center of each vortex. For every consecutive frame, the positions of each vortex center along with the corresponding magnitude were recorded. Due to the quasi-steady nature of the baseline wake, only a single set of tip vortices needed to be tracked. For the helix cases, a set of vortices was tracked after completion of each rotor revolution. 

When operating the turbine using greedy control, \Cref{fig:vortex_tracking} shows a very clear evolution of the tip vortices as they convect downstream. At around $x/D=1.75$, one can observe the leapfrogging of the first vortex pair. The first vortex is subsequently passed by the last trailing vortex, which completes the leapfrogging of the second vortex around $x/D=3$. While the vortices are seen to interact with each other, there is no clear sign of any instability yet with all the vortices staying more or less intact. The vorticity plots of the CCW helix case show one of the sets that experience the most aggresive mutual inductance. The leapfrogging of the first vortex pair has already taken place before $x/D=1.5$, which is earlier compared to the baseline case. Comparing the last frame of the two cases, we see that the third vortex has already jumped the first two vortices. On further inspection we can also notice that the set of vortices has convected less far downstream. This also shows how the helix influences the convection velocity of the vortices over time, which in turn causes sets of vortices to expand outward or contract towards the center of the wake. All previously mentioned phenomena are seen to destabilize the wake at a very fast rate.

\begin{figure}[hbt!]
    \centering
    \includegraphics[trim = {0, 0cm, 0, 0}, clip,scale=1]{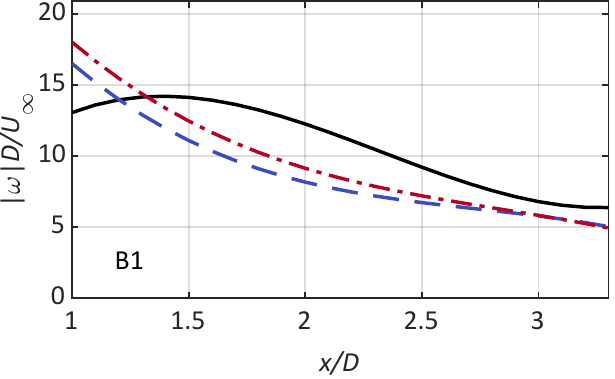}\hfill
    \includegraphics[trim = {0.67cm, 0cm, 0, 0}, clip,scale=1]{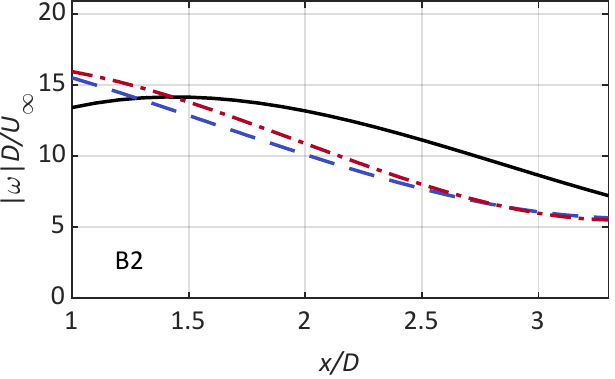}\hfill
    \includegraphics[trim = {0.67cm, 0cm, 0, 0}, clip,scale=1]{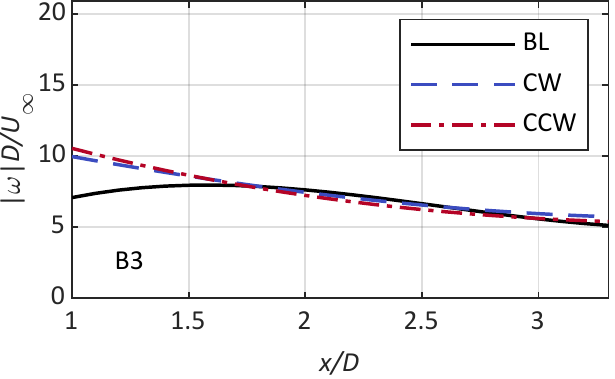}
    \caption{Vortex strength expressed by the normalized vorticity magnitude as a function of the displacement in the streamwise direction for three different test cases. Each figure corresponds to a vortex shed by one of the blades. The measurements were fitted to third order polynomial in order to obtain the average vorticity magnitude for each test case.}
    \label{fig:vortex_mag}
\end{figure}

One of the responsible factors for the higher mutual inductance of the tip vortices is the strength of the vortices (i.e., stronger vortices will have a larger relative influence, and thus increased interaction). By tracking all sets of vortices over time, we can also see how the strength of each vortex evolves over time (and distance downstream). Figure~\ref{fig:vortex_mag} presents the average vorticity magnitude of each trailing vortex as it travels downstream. The resulting plots were obtained by fitting a third order polynomial on all of the recorded vortex measurements. It is seen how the vorticity magnitude of both helix implementations is initially higher than the baseline case, leading to a stronger interaction initially. However, after the initial interaction between vortices, we observe that the magnitude decreases at a faster rate for the helix. Considering the magnitude of the third vortex, we see a striking difference compared to the first two vortices. For all three test cases, the initial vorticity is almost twice as low compared to the first and second vortices. Although the exact reason behind this difference is unknown, we reckon this is either the result of a slight offset in pitch angle of one of the blades, or imperfections on the blade surface near the tip. 

\begin{SCfigure}[][bt!]
\centering
    \includegraphics[scale=1]{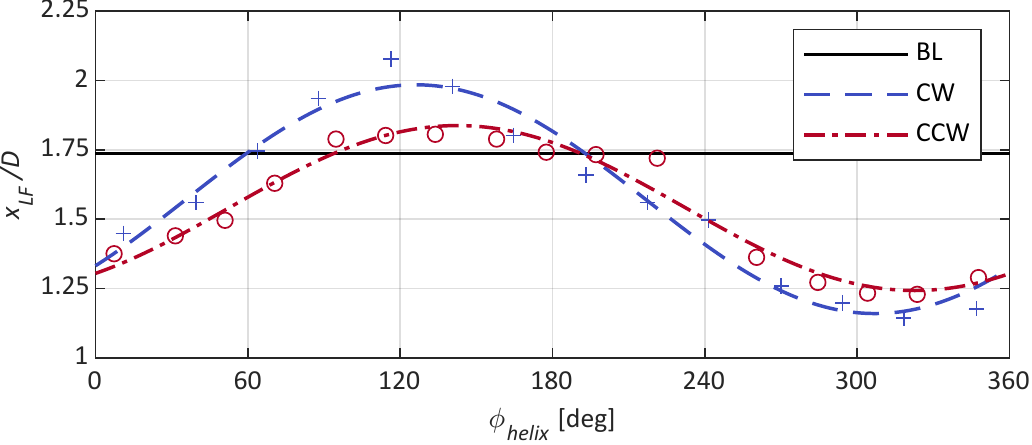}
    \caption{Comparison of the leapfrogging location $x_\text{LF}/D$ in the vertical plane ($y/D=0$) as a function of the helix azimuth $\phi_\text{helix}$, which is a measure of the position of the combined yaw and tilt moment on the rotor plane. A constant $x_\text{LF}/D$ is shown for the baseline case, while the varying locations for the helix approach are marked by $+$ and $\circ$ for the CW and CCW cases, respectively. These data points were subsequently fitted to a sine function of the form $a_1+a_2\sin(a_3t+a_4)$, with average offset $a_1$, amplitude $a_2$, frequency $a_3$ and phase offset $a_4$.}
    \label{fig:leapfrogging_location}
\end{SCfigure}

The tracking measurements also allow us to express the variation in the leapfrogging location of the first vortex pair as a function of the helix azimuth $\phi_\text{helix}$. For each of the vortex pairs, this location was marked when the second vortex moves over \SI{90}{\degree} with respect to the first vortex (i.e., $x_{v_2}(\phi)$ > $x_{v_1}(\phi)$, with $x_{v_i}$ indicating the streamwise location of a vortex $i$ for phase $\phi$). The resulting locations are presented in \Cref{fig:leapfrogging_location}, and compared to the constant leapfrogging location of the baseline case. The respective data points were subsequently fitted to a sine function, which shows how the pairing of the first two vortices evolves as the resultant yaw and tilt moment moves around the rotor plane. It appears that the CW helix has a slightly higher amplitude than the CCW case, probably related to the difference in pitch amplitude between the two cases. Furthermore, we show that on average the helix implementation initializes the leapfrogging at an earlier stage.

\subsection{Turbine measurements}
Besides analyzing the performance of the helix approach based on the flow fields, a more straightforward approach is to investigate the measurements of the turbines themselves. During each PIV recording, the generator power and bending moment at the tower bottom were measured. For each test case, all these measurements were processed to obtain the average and standard deviations. Since each PIV measurements was triggered at the same specific time instant, the periodic components in the generator power and bending moment were maintained, at least for turbine T1. 

\begin{figure}[t]
    \centering
    \includegraphics[scale=1]{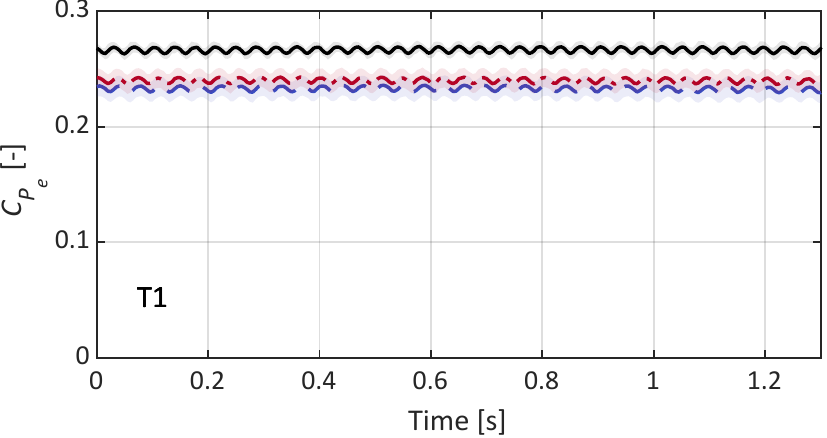}
    \hfill
    \includegraphics[scale=1]{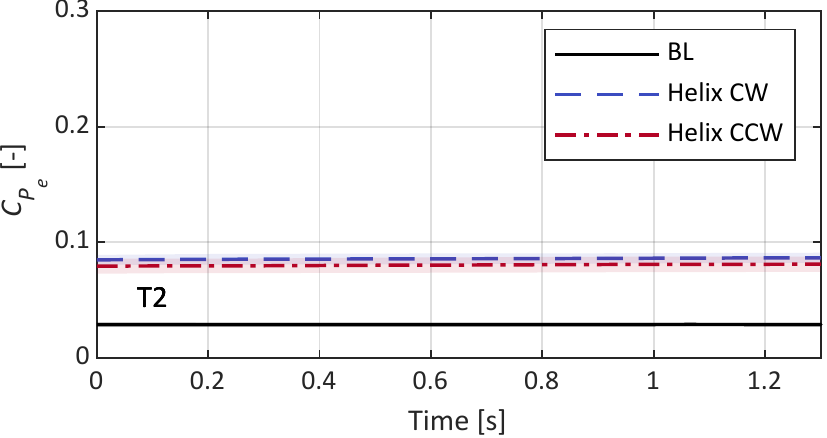}
    \caption{Sample of the average measured generator power of turbines T1 and T2. Some variation between different measurements of the same test case was observed. This variation is given by the shaded areas around the mean, indicating $\pm2\sigma$.}
    \label{fig:turbine_generator_power}
\end{figure}

The average generator power of both turbines is shown in \Cref{fig:turbine_generator_power} for the three test cases. The shaded areas represent the 95\% confidence intervals of the measurement variance. We observe that the baseline case is operating at the expected power level, while a significant decrease in power is observed for both helix cases. Compared to the power loss previously reported in simulations \citep{Frederik2020TheFarms,Taschner2023}, this decrease is quite severe. The larger drop in power is seen for the CW helix, probably due to the higher pitching amplitude. Measurements from the second turbine show that the helix approach is able to increase the power downstream by a large amount. In both cases, the original amount of power from the baseline case is more than doubled. Another observation is the periodic component present in all power measurements of T1. Closer inspection of the signal showed this as a $2P$ excitation, meaning it corresponds to two times the rotational frequency of the turbine. This $2P$ excitation in the fixed frame can be caused by an offset in pitch angle of one of the blades \citep{VanSolingen2015LinearTurbines}. This was already speculated on in the previous section when observing the vorticity magnitude of one of the tip vortices.

\begin{figure}[bt!]
\centering
    \includegraphics[scale=1]{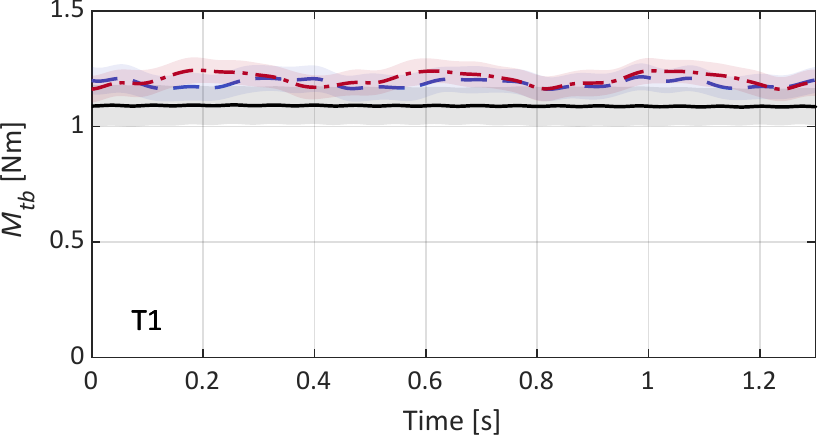}
    \hfill
    \includegraphics[scale=1]{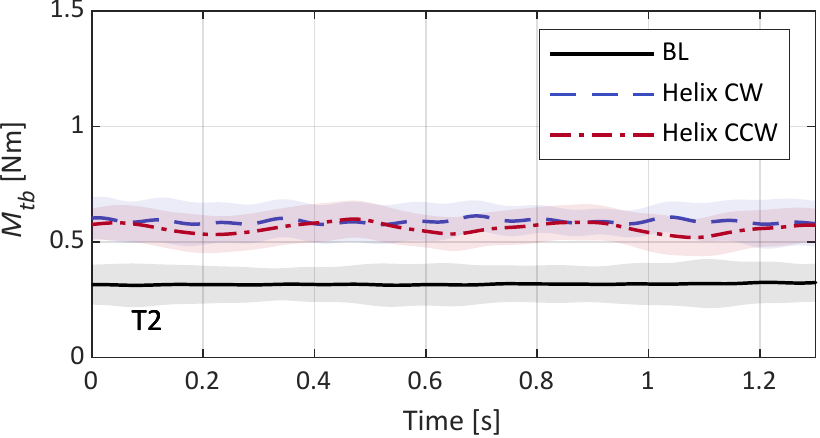}
     \caption{Sample of the average measured bending moment at the tower base of turbines T1 and T2. Some variation between different measurements of the same test case was observed. This variation is given by the shaded areas around the mean, indicating $\pm2\sigma$.}
    \label{fig:turbine_bending_moment}
\end{figure}

Figure~\ref{fig:turbine_bending_moment} provides the bending moments of both turbines. The periodic excitation of the tilt moment initiated by the helix approach is clearly visible, especially for the CCW helix. In both cases, the average bending moment experienced at the base of the tower increased. The bending moment of the second turbine shows a similar response. Again, the CCW helix especially shows the effect of the periodic tilt moment very clearly. While the average bending moment experienced by the downstream turbine is smaller compared to the first turbine, the relative increase is more severe for the second turbine. Both figures make it apparent, that while large gains in power are possible with the helix implementation, this will come at the cost of increased fatigue loading of some structural components. 

\begin{SCfigure}[][b]
\centering
    \includegraphics[scale=1]{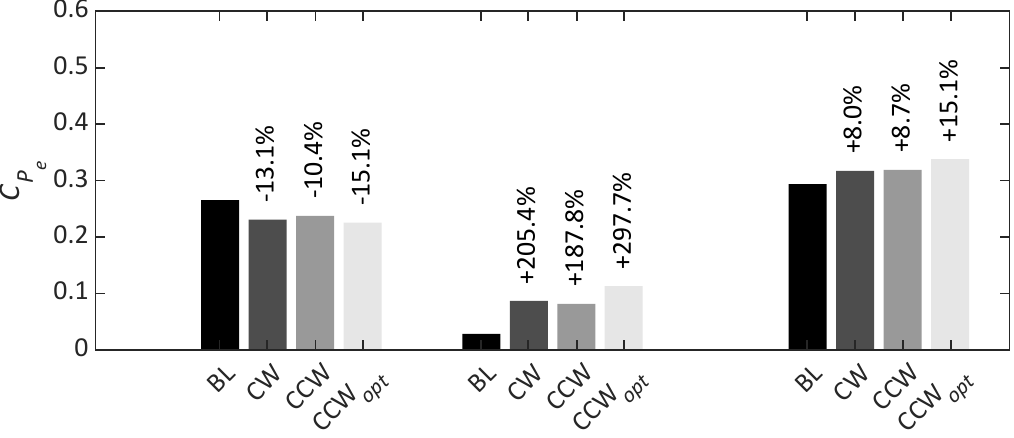}
    \caption{Average power production of turbines T1, T2, and the combined sum of both turbines for the different control strategies expressed by the electrical power coefficient $C_{P_e}$.}
    \label{fig:average_power_production}
\end{SCfigure}

Finally, the performance of maximizing the power output for all considered test cases, including the $\text{CCW}_\text{opt}$ implementation, is summarized in \Cref{fig:average_power_production}. The figure presents the average power generated by each turbine, as well as the combined power of T1 and T2. 
Due to the periodic pitching of the blades, T1 loses some aerodynamic efficiency, decreasing the power by more than 10\% for each implementation of the helix approach. The largest drop in power is observed for the $\text{CCW}_\text{opt}$ helix, indicating that the loss in performance can not only be ascribed to the pitch amplitude. The power of the second turbine was seen to double for the first two helix approaches, and even triple for the $\text{CCW}_\text{opt}$ helix. Adding up the power of both turbines, we observe that for this particular setup ($5\,D$ spacing with low turbulence), the helix approach was able to increase power by 8.0\% for the CW direction, by 8.7\% for the CCW direction, and by 15.1\% using the CCW helix with higher pitch amplitude. Based on the work of \citep{Taschner2023}, we believe even higher gains could be possible when larger pitch amplitudes are applied in combination with the CCW helix. 

\section{Conclusions}
This paper presented the implementation of the \emph{helix approach}, a promising control method for maximizing the power output of a wind farm, in a wind tunnel study. The experimental setup consisted of two scaled wind turbines ($D=\SI{0.58}{\meter})$ spaced $5\,D$ apart in front of an open jet. The upstream turbine was modified with a swashplate typically used for helicopter controls, allowing the turbine to mimick dynamic individual pitch control. A tomographic particle image velocimetry setup, using helium-filled soap bubbles as flow tracers, was used to measure the effect of the helix approach on the turbine wake. A comparison was made with a baseline case, which consisted of running the first turbine at the optimal power extraction settings (greedy control). Turbine measurements of the generator power and tower bending moment were collected as well.

The helix approach uses dynamic individual pitch control to create yaw and tilt moments from the rotor plane that are acting on the incoming flow. The resultant moment is then periodically moved over the rotor plane, resulting in a helical velocity profile in the wake. Two types of implementation are possible with the helix, either moving the helical velocity profile in clockwise or counter-clockwise direction. Previous work on the helix approach was mostly limited to simulation studies. This paper contributes to the experimental validation of the control method, which in time may lead to implementation of this method in the field. Using phase-averaged measurements of the streamwise velocity component, the amount of available power in the wake was determined for all test cases.  All helix implementations that were considered showed an earlier and enhanced recovery of the wake, providing more energy for turbines downstream. 

Phase-averaged flow fields were also used to investigate the wake recovery mechanism in further detail. Previous studies on the helix approach, as well as dynamic induction control with collective pitch, indicated that the performance of the control framework is related to the leapfrogging or vortex pairing instability of the shed tip vortices. It has been shown that this pairing initializes the breakdown of the tip vortices and is the onset for the re-energization of the wake. The periodic motion of the combined yaw and tilt moment results in tip vortices varying in strength and size, as well high induction zones being released into the wake. Therefore, the mutual inductance between vortices will also differ depending on time and place. Overall, it is believed these local variations will trigger the breakdown of vortices and destabilization of the wake, which will subsequently result in a forced meandering of the wake. The phase-averaged measurements showed a clear increase in the entrainment of kinetic energy in the wake due to random fluctuations.  Furthermore, after tracking the tip vortices over time using the vorticity magnitude, we showed that the average leapfrogging location of the first vortex pair is shortened with the helix approach.  

Finally, measurements from both turbines were used to corroborate the enhanced performance of the helix approach compared to normal operation, that was obtained through the flow measurements. The turbine that was used to implement the helix approach did suffer from a loss in aerodynamic efficiency, leading to decreases in power of over 10\%. However, these losses were more than made up for by the downstream turbine, enhancing the total power output of the two turbines by as much as 15\%. These results are even more promising when one considers that the losses at the first turbine are expected to decrease in case of utility scale wind turbines. The measurement results also confirm once more, that the CCW helix implementation is the more effective of the two.

\section*{acknowledgements}
This work is part of the Hollandse Kust Noord wind farm innovation program where CrossWind C.V., Shell, Eneco and Siemens Gamesa are teaming up; funding for the PhDs was provided by CrossWind C.V. and Siemens Gamesa.

We would like to extend great thanks to Will van Geest, Wim Wien, and the people from DEMO at TU Delft for their support in preparing the scaled wind turbine models. 

\section*{conflict of interest}
The authors declare no conflict of interest.

\section*{Supporting Information}
The experimental data that was presented in this paper has been uploaded to the 4TU research data base under doi 10.4121/22294423. 

\printendnotes

% Submissions are not required to reflect the precise reference formatting of the journal (use of italics, bold etc.), however it is important that all key elements of each reference are included.
\bibliography{references.bib,sample.bib}

\begin{thebibliography}{48}
\expandafter\ifx\csname natexlab\endcsname\relax\def\natexlab#1{#1}\fi
\expandafter\ifx\csname url\endcsname\relax
  \def\url#1{\texttt{#1}}\fi
\expandafter\ifx\csname urlprefix\endcsname\relax\def\urlprefix{URL: }\fi

\bibitem[{Aho et~al.(2012)Aho, Buckspan, Laks, Fleming, Jeong, Dunne,
  Churchfield, Pao and Johnson}]{Aho2012}
Aho, J., Buckspan, A., Laks, J., Fleming, P.~A., Jeong, Y., Dunne, F.,
  Churchfield, M., Pao, L.~Y. and Johnson, K. (2012) A tutorial of wind turbine
  control for supporting grid frequency through active power control.
\newblock \textit{American Control Conference (ACC)}, 3120--3131.

\bibitem[{Annoni et~al.(2016)Annoni, Gebraad, Scholbrock, Fleming and van
  Wingerden}]{Annoni2016a}
Annoni, J., Gebraad, P. M.~O., Scholbrock, A.~K., Fleming, P.~A. and van
  Wingerden, J.~W. (2016) Analysis of axial-induction-based wind plant control
  using an engineering and a high-order wind plant model.
\newblock \textit{Wind Energy}, \textbf{19}, 1135--1150.

\bibitem[{Barthelmie et~al.(2009)Barthelmie, Hansen, Frandsen, Rathmann,
  Schlez, Phillips, Hassan, Rados and Zervos}]{Barthelmie2009ModellingOffshore}
Barthelmie, R.~J., Hansen, K., Frandsen, S.~T., Rathmann, O., Schlez, W.,
  Phillips, J., Hassan, G., Rados, U.~K. and Zervos, A. (2009) {Modelling and
  Measuring Flow and Wind Turbine Wakes in Large Wind Farms Offshore}.

\bibitem[{Barthelmie et~al.(2013)Barthelmie, Hansen and
  Pryor}]{Barthelmie2013MeteorologicalWakes}
Barthelmie, R.~J., Hansen, K.~S. and Pryor, S.~C. (2013) {Meteorological
  controls on wind turbine wakes}.
\newblock \textit{Proceedings of the IEEE}, \textbf{101}, 1010--1019.

\bibitem[{Bir(2008)}]{Bir2008Multi-bladeAnalysis}
Bir, G. (2008) {Multi-blade coordinate transformation and its application to
  wind turbine analysis}.
\newblock \textit{46th AIAA Aerospace Sciences Meeting and Exhibit}.

\bibitem[{Bossanyi(2000)}]{Bossanyi2000TheTurbines}
Bossanyi, E.~A. (2000) {The Design of closed loop controllers for wind
  turbines}.
\newblock \textit{Wind Energy}, \textbf{3}, 149--163.

\bibitem[{Brown et~al.(2021)Brown, Houck, Maniaci, Westergaard and
  Kelley}]{Brown2021AcceleratedInstability}
Brown, K., Houck, D., Maniaci, D., Westergaard, C. and Kelley, C. (2021)
  {Accelerated Wind-Turbine Wake Recovery Through Actuation of the Tip-Vortex
  Instability}.

\bibitem[{Cal et~al.(2010)Cal, Lebr{\'{o}}n, Castillo, Kang and
  Meneveau}]{Cal2010ExperimentalLayer}
Cal, R.~B., Lebr{\'{o}}n, J., Castillo, L., Kang, H.~S. and Meneveau, C. (2010)
  {Experimental study of the horizontally averaged flow structure in a model
  wind-turbine array boundary layer}.
\newblock \textit{Journal of Renewable and Sustainable Energy}, \textbf{2},
  013106.

\bibitem[{Campagnolo et~al.(2016)Campagnolo, Petrović, Bottasso and
  Croce}]{Campagnolo2016}
Campagnolo, F., Petrović, V., Bottasso, C.~L. and Croce, A. (2016) Wind tunnel
  testing of wake control strategies.
\newblock \textit{Proceedings of the American Control Conference (ACC)},
  513--518.

\bibitem[{Coquelet(2022)}]{CoqueletThesis}
Coquelet, M. (2022) \textit{Numerical investigation of wind turbine control
  schemes for load alleviation and wake effects mitigation}.
\newblock Ph.D. thesis, Université catholique de Louvain.

\bibitem[{Croce et~al.(2022)Croce, Cacciola, Montero Montenegro, Stipa and
  Pratic{\'{o}}}]{Croce2022AApplications}
Croce, A., Cacciola, S., Montero Montenegro, M., Stipa, S. and Pratic{\'{o}},
  R. (2022) {A CFD-based analysis of dynamic induction techniques for wind farm
  control applications}.
\newblock \textit{Wind Energy}.

\bibitem[{Doekemeijer et~al.(2021)Doekemeijer, Kern, Maturu, Kanev, Salbert,
  Schreiber, Campagnolo, Bottasso, Schuler, Wilts, Neumann, Potenza,
  Calabretta, Fioretti and van Wingerden}]{Doekemeijer2021FieldItaly}
Doekemeijer, B.~M., Kern, S., Maturu, S., Kanev, S., Salbert, B., Schreiber,
  J., Campagnolo, F., Bottasso, C.~L., Schuler, S., Wilts, F., Neumann, T.,
  Potenza, G., Calabretta, F., Fioretti, F. and van Wingerden, J.~W. (2021)
  {Field experiment for open-loop yaw-based wake steering at a commercial
  onshore wind farm in Italy}.
\newblock \textit{Wind Energy Science}, \textbf{6}, 159--176.

\bibitem[{Fleming et~al.(2019)Fleming, King, Dykes, Simley, Roadman,
  Scholbrock, Murphy, K.~Lundquist, Moriarty, Fleming, van Dam, Bay, Mudafort,
  Lopez, Skopek, Scott, Ryan, Guernsey and Brake}]{Fleming2019}
Fleming, P., King, J., Dykes, K., Simley, E., Roadman, J., Scholbrock, A.,
  Murphy, P., K.~Lundquist, J., Moriarty, P., Fleming, K., van Dam, J., Bay,
  C., Mudafort, R., Lopez, H., Skopek, J., Scott, M., Ryan, B., Guernsey, C.
  and Brake, D. (2019) Initial results from a field campaign of wake steering
  applied at a commercial wind farm: Part 1.
\newblock \textit{Wind Energy Science Discussions}, 1--22.

\bibitem[{Fleming et~al.(2016{\natexlab{a}})Fleming, Aho, Buckspan, Ela, Zhang,
  Gevorgian, Scholbrock, Pao and Damiani}]{Fleming2016EffectsLoading}
Fleming, P.~A., Aho, J., Buckspan, A., Ela, E., Zhang, Y., Gevorgian, V.,
  Scholbrock, A., Pao, L. and Damiani, R. (2016{\natexlab{a}}) {Effects of
  power reserve control on wind turbine structural loading}.
\newblock \textit{Wind Energy}, \textbf{19}, 453--469.

\bibitem[{Fleming et~al.(2016{\natexlab{b}})Fleming, Aho, Gebraad, Pao and
  Zhang}]{Fleming2016}
Fleming, P.~A., Aho, J., Gebraad, P. M.~O., Pao, L.~Y. and Zhang, Y.
  (2016{\natexlab{b}}) Computational fluid dynamics simulation study of active
  power control in wind plants.
\newblock \textit{American Control Conference (ACC)}, 1413--1420.

\bibitem[{Fleming et~al.(2017)Fleming, Annoni, Scholbrock, Quon, Dana, Schreck,
  Raach, Haizmann and Schlipf}]{Fleming2017b}
Fleming, P.~A., Annoni, J., Scholbrock, A.~K., Quon, E., Dana, S., Schreck, S.,
  Raach, S., Haizmann, F. and Schlipf, D. (2017) Full-scale field test of wake
  steering.
\newblock \textit{Journal of Physics: Conference Series}, \textbf{854}.

\bibitem[{Frederik et~al.(2020{\natexlab{a}})Frederik, Doekemeijer, Mulders and
  van Wingerden}]{Frederik2020TheFarms}
Frederik, J.~A., Doekemeijer, B.~M., Mulders, S.~P. and van Wingerden, J.-W.
  (2020{\natexlab{a}}) {The helix approach: using dynamic individual pitch
  control to enhance wake mixing in wind farms}.
\newblock \textit{Wind Energy}.

\bibitem[{Frederik et~al.(2020{\natexlab{b}})Frederik, Weber, Cacciola,
  Campagnolo, Croce, Bottasso and van
  Wingerden}]{Frederik2020PeriodicExperiments}
Frederik, J.~A., Weber, R., Cacciola, S., Campagnolo, F., Croce, A., Bottasso,
  C. and van Wingerden, J.-W. (2020{\natexlab{b}}) {Periodic dynamic induction
  control of wind farms: proving the potential in simulations and wind tunnel
  experiments}.
\newblock \textit{Wind Energy Science}, \textbf{5}, 245--257.

\bibitem[{Frederik and van Wingerden(2022)}]{Frederik2022OnStrategies}
Frederik, J.~A. and van Wingerden, J.~W. (2022) {On the load impact of dynamic
  wind farm wake mixing strategies}.
\newblock \textit{Renewable Energy}, \textbf{194}, 582--595.

\bibitem[{Gebraad et~al.(2016)Gebraad, Teeuwisse, van Wingerden, Fleming,
  Ruben, Marden and Pao}]{Gebraad2016}
Gebraad, P. M.~O., Teeuwisse, F.~W., van Wingerden, J.~W., Fleming, P.~A.,
  Ruben, S.~D., Marden, J.~R. and Pao, L.~Y. (2016) Wind plant power
  optimization through yaw control using a parametric model for wake effects -
  a cfd simulation study.
\newblock \textit{Wind Energy}, \textbf{19}, 95--114.

\bibitem[{Goit and Meyers(2015)}]{Meyers2015}
Goit, J.~P. and Meyers, J. (2015) Optimal control of energy extraction in
  wind-farm boundary layers.
\newblock \textit{Journal of Fluid Mechanics}, \textbf{768}, 5--50.

\bibitem[{Heckmeier(2022)}]{FlorianHeckmeier}
Heckmeier, F.~M. (2022) \textit{Multi-Hole Probes for Unsteady Aerodynamics
  Analysis}.
\newblock Ph.D. thesis, Technische Universität München.

\bibitem[{van~der Hoek et~al.(2022)van~der Hoek, Frederik, Huang, Scarano,
  Simao~Ferreira and van Wingerden}]{VanDerHoek2022ExperimentalWake}
van~der Hoek, D., Frederik, J., Huang, M., Scarano, F., Simao~Ferreira, C. and
  van Wingerden, J.-W. (2022) {Experimental analysis of the effect of dynamic
  induction control on a wind turbine wake}.
\newblock \textit{Wind Energy Science}, \textbf{7}, 1305--1320.

\bibitem[{van~der Hoek et~al.(2019)van~der Hoek, Kanev, Allin, Bieniek and
  Mittelmeier}]{Hoek2019}
van~der Hoek, D., Kanev, S., Allin, J., Bieniek, D. and Mittelmeier, N. (2019)
  Effects of axial induction control on wind farm energy production - a field
  test.
\newblock \textit{Renewable Energy}, \textbf{140}, 994--1003.

\bibitem[{van~der Hoek et~al.(2018)van~der Hoek, Kanev and
  Engels}]{VanDerHoek2018ComparisonLoads}
van~der Hoek, D., Kanev, S. and Engels, W. (2018) {Comparison of
  Down-Regulation Strategies for Wind Farm Control and their Effects on Fatigue
  Loads}.
\newblock In \textit{Proceedings of the American Control Conference}, vol.
  2018-June.

\bibitem[{Houck(2022)}]{Houck2022ReviewTurbines}
Houck, D.~R. (2022) {Review of wake management techniques for wind turbines}.
\newblock \textit{Wind Energy}, \textbf{25}, 195--220.

\bibitem[{Kanev et~al.(2018)Kanev, Savenije and Engels}]{Kanev2018}
Kanev, S., Savenije, F. and Engels, W. (2018) Active wake control: An approach
  to optimize the lifetime operation of wind farms.
\newblock \textit{Wind Energy}, \textbf{21}, 488--501.

\bibitem[{Korb et~al.(2021)Korb, Asmuth, Stender and
  Ivanell}]{Korb2021ExploringControl}
Korb, H., Asmuth, H., Stender, M. and Ivanell, S. (2021) {Exploring the
  application of reinforcement learning to wind farm control}.
\newblock \textit{Journal of Physics: Conference Series}, \textbf{1934},
  012022.

\bibitem[{Lignarolo et~al.(2015)Lignarolo, Ragni, Scarano, Sim{\~{a}}o~Ferreira
  and Van~Bussel}]{Lignarolo2015Tip-vortexWakes}
Lignarolo, L.~E., Ragni, D., Scarano, F., Sim{\~{a}}o~Ferreira, C.~J. and
  Van~Bussel, G.~J. (2015) {Tip-vortex instability and turbulent mixing in
  wind-turbine wakes}.
\newblock \textit{Journal of Fluid Mechanics}, \textbf{781}, 467--493.

\bibitem[{Lignarolo et~al.(2014)Lignarolo, Ragni, Ferreira and
  Van~Bussel}]{Lignarolo2014ExperimentalVelocimetry}
Lignarolo, L. E.~M., Ragni, D., Ferreira, C. J.~S. and Van~Bussel, G. J.~W.
  (2014) {Experimental quantification of the entrainment of kinetic energy and
  production of turbulence in the wake of a wind turbine with Particle Image
  Velocimetry}.

\bibitem[{Meyers et~al.(2022)Meyers, Bottasso, Dykes, Fleming, Gebraad, Giebel,
  G{\"{o}}{\c{c}}men and van Wingerden}]{Meyers2022WindChallenges}
Meyers, J., Bottasso, C., Dykes, K., Fleming, P., Gebraad, P., Giebel, G.,
  G{\"{o}}{\c{c}}men, T. and van Wingerden, J.-W. (2022) {Wind farm flow
  control: Prospects and challenges}.
\newblock \textit{Wind Energy Science}, \textbf{7}, 2271--2306.

\bibitem[{Munters and Meyers(2017)}]{Munters2017}
Munters, W. and Meyers, J. (2017) An optimal control framework for dynamic
  induction control of wind farms and their interaction with the atmospheric
  boundary layer.
\newblock \textit{Philosophical Transactions of the Royal Society of London A:
  Mathematical, Physical and Engineering Sciences}, \textbf{375}.

\bibitem[{Munters and Meyers(2018)}]{Munters2018TowardsTurbines}
--- (2018) {Towards practical dynamic induction control of wind farms: Analysis
  of optimally controlled wind-farm boundary layers and sinusoidal induction
  control of first-row turbines}.
\newblock \textit{Wind Energy Science}, \textbf{3}, 409--425.

\bibitem[{Muscari et~al.(2022)Muscari, Schito, Vir{\'{e}}, Zasso, van~der Hoek
  and van Wingerden}]{Muscari2022PhysicsFarms}
Muscari, C., Schito, P., Vir{\'{e}}, A., Zasso, A., van~der Hoek, D. and van
  Wingerden, J.~W. (2022) {Physics informed DMD for periodic Dynamic Induction
  Control of Wind Farms}.
\newblock \textit{Journal of Physics: Conference Series}, \textbf{2265},
  022057.

\bibitem[{Reynolds and Hussain(1972)}]{Reynolds1972TheExperiments}
Reynolds, W.~C. and Hussain, A.~K. (1972) {The mechanics of an organized wave
  in turbulent shear flow. Part 3. Theoretical models and comparisons with
  experiments}.
\newblock \textit{Journal of Fluid Mechanics}, \textbf{54}, 263--288.

\bibitem[{Scarano et~al.(2015)Scarano, Ghaemi, Caridi, Bosbach, Dierksheide and
  Sciacchitano}]{Scarano2015OnExperiments}
Scarano, F., Ghaemi, S., Caridi, G. C.~A., Bosbach, J., Dierksheide, U. and
  Sciacchitano, A. (2015) {On the use of helium-filled soap bubbles for
  large-scale tomographic PIV in wind tunnel experiments}.
\newblock \textit{Experiments in Fluids}, \textbf{56}.

\bibitem[{Schanz et~al.(2016)Schanz, Gesemann and
  Schr{\"{o}}der}]{Schanz2016Shake-The-Box:Densities}
Schanz, D., Gesemann, S. and Schr{\"{o}}der, A. (2016) {Shake-The-Box:
  Lagrangian particle tracking at high particle image densities}.
\newblock \textit{Experiments in Fluids}, \textbf{57}, 70.

\bibitem[{Schottler et~al.(2016)Schottler, H{\"{o}}lling, Peinke and
  H{\"{o}}lling}]{Schottler2016DesignStudies}
Schottler, J., H{\"{o}}lling, A., Peinke, J. and H{\"{o}}lling, M. (2016)
  {Design and implementation of a controllable model wind turbine for
  experimental studies}.
\newblock \textit{Journal of Physics: Conference Series}, \textbf{753}.

\bibitem[{Sciacchitano and Scarano(2014)}]{Sciacchitano2014EliminationFilter}
Sciacchitano, A. and Scarano, F. (2014) {Elimination of PIV light reflections
  via a temporal high pass filter}.
\newblock \textit{Measurement Science and Technology}, \textbf{25}.

\bibitem[{van Solingen and van Wingerden(2015)}]{VanSolingen2015LinearTurbines}
van Solingen, E. and van Wingerden, J.-W. (2015) {Linear individual pitch
  control design for two-bladed wind turbines}.
\newblock \textit{Wind Energy}, \textbf{18}, 677--697.

\bibitem[{S{\o}rensen(2011)}]{Srensen2011InstabilityWakes}
S{\o}rensen, J.~N. (2011) {Instability of helical tip vortices in rotor wakes}.
\newblock \textit{Journal of Fluid Mechanics}, \textbf{682}, 1--4.

\bibitem[{Taschner et~al.(2023)Taschner, van Vondelen, Verzijlbergh and van
  Wingerden}]{Taschner2023}
Taschner, E., van Vondelen, A., Verzijlbergh, R. and van Wingerden, J.-W.
  (2023) {On the performance of the helix wind farm control approach in the
  conventionally neutral atmospheric boundary layer}.
\newblock \textit{submitted to Wake Conference 2023}.

\bibitem[{Vali et~al.(2017)Vali, Petrovi\'c, Boersma, Van~Wingerden and
  K\"uhn}]{Vali2017}
Vali, M., Petrovi\'c, V., Boersma, S., Van~Wingerden, J.~W. and K\"uhn, M.
  (2017) Adjoint-based model predictive control of wind farms: Beyond the quasi
  steady-state power maximization.
\newblock \textit{IFAC World Congress}, \textbf{50}, 4510 -- 4515.

\bibitem[{Vollmer et~al.(2016)Vollmer, Steinfeld, Heinemann and
  K{\"{u}}hn}]{Vollmer2016EstimatingStudy}
Vollmer, L., Steinfeld, G., Heinemann, D. and K{\"{u}}hn, M. (2016) {Estimating
  the ake deflection donstream of a ind turbine in different atmospheric
  stabilities: An les study}.
\newblock \textit{Wind Energy Science}, \textbf{1}, 129--141.

\bibitem[{van Vondelen et~al.(2023)van Vondelen, Navalkar, Kerssemakers and van
  Wingerden}]{vanVondelen2023}
van Vondelen, A., Navalkar, S., Kerssemakers, D. and van Wingerden, J.-W.
  (2023) {Enhanced wake mixing in wind farms using the Helix approach: A loads
  sensitivity study}.
\newblock \textit{submitted to Wake Conference 2023}.

\bibitem[{Wieneke(2008)}]{Wieneke2008VolumeVelocimetry}
Wieneke, B. (2008) {Volume self-calibration for 3D particle image velocimetry}.
\newblock \textit{Experiments in Fluids 2008 45:4}, \textbf{45}, 549--556.

\bibitem[{van Wingerden et~al.(2020)van Wingerden, Fleming, G{\"{o}}{\c{c}}men,
  Eguinoa, Doekemeijer, Dykes, Lawson, Simley, King, Astrain, Iribas, Bottasso,
  Meyers, Raach, K{\"{o}}lle and Giebel}]{vanWingerden2020ExpertControl}
van Wingerden, J.~W., Fleming, P.~A., G{\"{o}}{\c{c}}men, T., Eguinoa, I.,
  Doekemeijer, B.~M., Dykes, K., Lawson, M., Simley, E., King, J., Astrain, D.,
  Iribas, M., Bottasso, C.~L., Meyers, J., Raach, S., K{\"{o}}lle, K. and
  Giebel, G. (2020) {Expert Elicitation on Wind Farm Control}.
\newblock \textit{Journal of Physics: Conference Series}, \textbf{1618},
  022025.

\bibitem[{Yılmaz and Meyers(2018)}]{Ylmaz2018OptimalTurbines}
Yılmaz, A.~E. and Meyers, J. (2018) {Optimal dynamic induction control of a
  pair of inline wind turbines}.
\newblock \textit{Physics of Fluids}, \textbf{30}.

\end{thebibliography}

% \begin{biography}[example-image-1x1]{A.~One}
% Please check with the journal's author guidelines whether author biographies are required. They are usually only included for review-type articles, and typically require photos and brief biographies (up to 75 words) for each author.
% \bigskip
% \bigskip
% \end{biography}

%\newpage
%\graphicalabstract{example-image-1x1}{Please check the journal's author guildines for whether a graphical abstract, key points, new findings, or other items are required for display in the Table of Contents.}

\end{document}